\documentclass[tightenlines,twoside,secnumarabic,
               onecolumn,floatfix,nofootinbib,showpacs,11pt]{revtex4}

\usepackage[english]{babel}
\usepackage{pifont}
\usepackage{amsmath,amssymb,slashed}
\usepackage{enumerate}
\usepackage{graphicx}
\usepackage[sort&compress]{natbib}
\usepackage{xcolor}
\usepackage[normalem]{ulem}
\definecolor{red}{rgb}{1.0, 0, 0}

\allowdisplaybreaks

\bibpunct{[}{]}{,}{n}{}{,}

\newcommand{\bra}[1]{\ensuremath{\langle #1 |}}   % Bra vector
\newcommand{\ket}[1]{\ensuremath{| #1 \rangle}}   % Ket vector
\newcommand{\ev}[1]{\ensuremath{\left\langle #1 %
                     \right\rangle}} % Expectation value
\newcommand{\iso}[2]{{\ensuremath{{}^{#2}}\ensuremath{\rm #1}}}
\newcommand{\eps}{{\ensuremath{\epsilon}}}

\newcommand{\parenbar}[1]{\overset{
            \raisebox{-0.15em}{\scalebox{.4}{\textbf{(}}}
            \raisebox{-0.1em}{{\hspace{.03em}\_\hspace{.05em}}}
            \raisebox{-0.15em}{\scalebox{.4}{\textbf{)}}}} {#1}}

\definecolor{dgreen}{rgb}{0,0.5,0}
\definecolor{dred}{rgb}{0.7,0,0}
\newcommand{\dgreen}{\color{dgreen}}
\newcommand{\dred}{\color{dred}}
\newcommand{\OK}{{\dgreen\ding{51}}}
\newcommand{\NO}{{\dred\ding{55}}}
\newcommand{\NOTKNOWN}{{\dred\bf ?}}

\begin{document}

% =============================================================================
\title{Exploring $\nu$ signals in dark matter detectors}
\author{Roni Harnik$^1$}             \email[Email: ]{roni@fnal.gov}
\author{Joachim Kopp$^1$}            \email[Email: ]{jkopp@fnal.gov}
\author{Pedro A.~N.~Machado$^{1,2,3}$} \email[Email: ]{accioly@fma.if.usp.br}
\affiliation{$^1$ Fermilab, P.O.~Box 500, Batavia, IL 60510-0500, USA \\
             $^2$ Instituto de F\'{i}sica, Universidade de S\~{a}o Paulo,
             C.P.~66.318, 05315-970 S\~{a}o Paulo, Brazil \\
             $^3$ Institut de Physique Th\'{e}orique, CEA Saclay,
             91191 Gif-sur-Yvette, France}            
\date{February 27, 2011} % FIXME
% =============================================================================

\begin{abstract}
We investigate standard and non-standard solar neutrino signals in direct dark
matter detection experiments.  It is well known that even without new physics,
scattering of solar neutrinos on nuclei or electrons is an irreducible
background for direct dark matter searches, once these experiments reach the
ton scale.  Here, we entertain the possibility that neutrino interactions are
enhanced by new physics, such as new light force carriers (for instance a
``dark photon'') or neutrino magnetic moments. We consider models with only the
three standard neutrino flavors, as well as scenarios with extra sterile
neutrinos.  We find that low-energy neutrino--electron and neutrino--nucleus
scattering rates can be enhanced by several orders of magnitude, potentially
enough to explain the event excesses observed in CoGeNT and CRESST. We also
investigate temporal modulation in these neutrino signals, which can arise from
geometric effects, oscillation physics, non-standard neutrino energy loss, and
direction-dependent detection efficiencies. We emphasize that, in addition to
providing potential explanations for existing signals, models featuring new
physics in the neutrino sector can also be very relevant to future dark matter
searches, where, on the one hand, they can be probed and constrained, but on
the other hand, their signatures could also be confused with dark matter
signals.
\end{abstract}

%==============================================================================
\begin{flushright}
  FERMILAB-PUB-12-048-T
\end{flushright}

\maketitle
%==============================================================================

%==============================================================================
\section{Introduction and motivation}
%==============================================================================

Experiments searching for dark matter and those studying solar neutrinos have
coexisted for many years, and even though many of the technological
challenges---such as the suppression of radioactive backgrounds and the
lowering of the energy threshold---are similar for both types of experiments,
their physics programs have had little overlap. In solar neutrino physics, the
size of the detector is more important than the energy threshold, whereas the
direct search for dark matter scattering is only possible for low energy
thresholds of order 10~keV or below. Until very recently, achieving such a low
threshold has only been possible in small detectors with masses of only a few
kg, too small to be of interest to neutrino physics. However, as dark matter
detectors are becoming more and more massive, this is about to change. In ton
scale experiments, solar neutrinos will constitute a non-negligible irreducible
background~\cite{Gutlein:2010tq} which can only be overcome in detectors with
directional sensitivity.

In this work we consider the additional possibility that there is new physics
in the neutrino sector, and we will examine how this may affect the signals
observed by present and future dark matter experiments. In particular, we will
discuss models featuring a new light ($\ll 1$~GeV) gauge boson $A'$. We will
also discuss the possible existence of neutrino magnetic moments and of extra
(``sterile'') neutrino species coupled to the new gauge boson.\footnote{In a
slight abuse of nomenclature, we will call the extra neutrino flavors
``sterile'' even though they can interact with ordinary matter through $A'$
exchange.} Sterile neutrinos are motivated by a number of anomalous results
from short-baseline neutrino oscillation experiments~\cite{Aguilar:2001ty,
AguilarArevalo:2010wv, Mueller:2011nm, Huber:2011wv, Mention:2011rk,
Karagiorgi:2009nb, Kopp:2011qd, Giunti:2011gz, Giunti:2011hn,
Karagiorgi:2011ut} and by cosmology~\cite{Krauss:2010xg, Giusarma:2011ex,
Hamann:2011ge}, whereas new light force carriers appear for instance in models
of Sommerfeld-enhanced dark matter annihilation (see
e.g.~\cite{ArkaniHamed:2008qn, Meade:2009rb, Fox:2008kb, Cheung:2009qd,
Katz:2009qq}). Both sterile neutrinos and light force carriers were used by
Nelson and Walsh to explain some of the neutrino oscillations
anomalies~\cite{Nelson:2007yq, Engelhardt:2010dx} (see
also~\cite{Karagiorgi:2012kw} for related work).  As also pointed out by
Pospelov~\cite{Pospelov:2011ha}, a new light force carrier can enhance the
interaction rates of active and sterile neutrinos at low energy by several
orders of magnitude compared to Standard Model neutrinos. Additionally, it can
have a large impact on the Mikheyev-Smirnov-Wolfenstein (MSW) potential that
neutrinos feel when propagating through matter, thus affecting oscillations.

As examples, we will consider potential explanations of current dark matter
anomalies, in particular the event excesses observed by
CoGeNT~\cite{Aalseth:2010vx, Aalseth:2011wp} and CRESST~\cite{Angloher:2011uu},
as well as the annual modulation signals observed in
CoGeNT~\cite{Aalseth:2011wp} and DAMA~\cite{Bernabei:2010mq}. While we will see
that finding a model that is consistent with all current direct detection data
and with constraints from neutrino oscillation experiments is difficult (a fate
that our scenario shares with most dark matter models), we will also see that
future experiments can be considerably affected by new physics in the neutrino
sector. In particular, neutrino signals beyond from physics beyond the Standard
Model may be confused with genuine dark matter signals. (The problem of other
types of new physics faking a dark matter signal is well-known in the context of
collider searches~\cite{ArkaniHamed:1998rs, Rizzo:2008fp, Drees:2012dd, Friedland:2011za}
but has received less attention in the framework of direct searches.)

It will be important for us that dedicated neutrino experiments have a
much higher energy threshold than dark matter detectors, which can detect
electron recoil energies as low as 0.5~keVee and nuclear recoil energies down
to 2~keVnr\footnote{The notation keVee refers to
the ``electron equivalent energy in keV'', which is defined as the
reconstructed recoil energy under the assumption that it is carried by an
electron. For nuclear recoils, only part of the recoil energy is visible in the
detector---an effect which has to be corrected for by dividing the visible
energy by a quenching factor---so that the energy threshold for nuclear recoils
is higher than that for electron recoils. When referring to a nuclear recoil
energy, we will use the notation ``keVnr''.} 
 in the case of CoGeNT~\cite{Aalseth:2010vx} and
CDMS~\cite{Ahmed:2010wy}. Thus, if new physics in the neutrino sector manifests
itself most strongly at low energy, neutrino detectors would be insensitive,
while dark matter detectors have excellent prospects of discovering or
constraining such models.

The outline of the paper is as follows: In section~\ref{sec:rates}, we review
neutrino scattering in the Standard Model and compute the expected
neutrino--electron and neutrino--nucleus scattering rates in dark matter
detectors. We then introduce in section~\ref{sec:model-intro} four
representative extension of the Standard Model that can lead to enhanced
neutrino interactions at low energy. In particular, we will discuss neutrino
magnetic moments, light new gauge bosons with various coupling structures, and
sterile neutrinos.  The phenomenology of these models is then analyzed in
section~\ref{sec:electron-recoil} with regard to neutrino--electron scattering
and in section~\ref{sec:nuclear-recoil} with regard to neutrino--nucleus
scattering.  In section~\ref{sec:modulation}, we discuss several mechanisms that
can lead to diurnal, semi-annual and annual modulation of the neutrino count
rate at the Earth: The varying Earth--Sun distance, oscillation with very long
($\sim 1$~AU) oscillation length, Earth matter effects, sterile neutrino
absorption, and direction-dependent detection efficiencies. Finally, in
section~\ref{sec:paramspace}, we discuss existing experimental limits on light
gauge bosons, and we show how they constrain the models introduced earlier.  We
also use Borexino and GEMMA data to set new limits on models with light force
carriers.  We summarize and conclude in section~\ref{sec:conclusions}.

%==============================================================================
\section{Neutrino Interactions in Dark Matter Detectors in the Standard Model}
\label{sec:rates}
%==============================================================================

Solar neutrinos may scatter elastically with nuclei or with electrons in a
target and produce low energy recoil events. Within the Standard Model~\cite{Fox:2005yp}, the
interaction rates are very low, beyond the reach of current and near future
dark matter detectors, but well within the reach of the larger dedicated solar
neutrino experiments Borexino~\cite{Borexino:2011rx} and
SNO~\cite{Aharmim:2005gt, Aharmim:2009gd}.  Solar neutrino interactions will
also become relevant to dark matter experiments once these experiments reach a
sensitivity to dark matter--nucleon scattering cross sections of order
$10^{-46}$~cm$^2$~\cite{Gutlein:2010tq}. 

The differential cross section for neutrino--electron scattering in the
Standard Model is easily found  to be
\begin{align}
  \frac{d\sigma_{\rm SM}(\nu_e e \to \nu_e e)}{dE_r} &=
  \frac{G_F^2 m_e}{2 \pi  E_\nu^2}
    \Big[ 4 s_w^4 (2 E_\nu^2+E_r^2-E_r (2 E_\nu+m_e))-2s_w^2 (E_r m_e-2 E_\nu^2)+
        E_\nu^2 \Big] \,,
  \label{eq:dsigmadE-nue-e} \\
\frac{d\sigma_{\rm SM}(\nu_{\mu,\tau} e \to \nu_{\mu,\tau} e)}{dE_r}  &=
  \frac{G_F^2 m_e}{2 \pi E_\nu^2}
    \Big[ 4 s_w^4 (2 E_\nu^2+E_r^2-E_r (2 E_\nu+m_e))+2s_w^2 (E_r m_e-2 E_\nu^2)+
        E_\nu^2 \Big] \,
  \label{eq:dsigmadE-numu-e}
\end{align}
for electron neutrinos ($\nu_e$) and muon/tau neutrinos ($\nu_\mu$,
$\nu_\tau$), respectively.  In these expressions, $G_F$ is the Fermi constant,
$s_w$ is the sine of the Weinberg angle, $E_\nu$ and $E_r$ are the neutrino
energy and the recoil energy transferred to the target electron, respectively,
and $m_e$ is the electron mass.  Note that scattering of electron neutrinos on
electrons receives contributions from $s$-channel $W$ exchange and from
$t$-channel $Z$ exchange, whereas for muon and tau neutrinos, only $Z$ exchange
is possible.  The differential cross section for neutrino--nucleus scattering
is
\begin{align}
  \frac{d\sigma_{\rm SM}(\nu_{e,\mu,\tau} N \to \nu_{e,\mu,\tau} N)}{dE_r} &=
  \frac{G_F^2 m_N F^2(E_r)}{4 \pi} \left[Z(4s_w^2-1)+N\right]^2
                                          \left(1-\frac{m_N E_r}{2E_\nu^2}\right)
  \label{eq:dsigmadE-nu-N}
\end{align}
for a nucleus of mass $m_N$, neutron number $N$ and charge $Z$. The
function $F(E_r)$ is the nuclear form factor, for which we will use
the form $F(E_r) = 3 e^{-\kappa^2 s^2/2} [\sin(\kappa r)-\kappa
  r\cos(\kappa r)] / (\kappa r)^3$, with $s = 1$~fm, $r = \sqrt{R^2 -
  5 s^2}$, $R = 1.2 A^{1/3}$~fm, $\kappa = \sqrt{2 m_N E_r}$ (and
$q^2\simeq-\kappa^2$)~\cite{Engel:1991wq}.  Terms proportional to
$E_\nu^2/m_N^2$ or $E_r/m_N$ were dropped. Note that for heavy nuclei
low-energy neutrino--nucleus scattering is enhanced compared to
scattering on light nuclei because the cross section contains terms
proportional to $N^2$, $Z^2$ and $N Z$. These terms reflect the fact
that neutrinos scatter coherently off all the nucleons. At higher
recoil energies, the form factor becomes relevant and compensates part
or all of this enhancement.

The maximum recoil energy of a target particle of mass $m_T$ for fixed $E_\nu$ is given by
\begin{align}
  E_r^{\mathrm{max}} = \frac{2 E_\nu^2}{m_T + 2 E_\nu }\,,
\end{align} 
and, conversely, the minimum neutrino energy required for transferring a given
recoil energy $E_r$ is
\begin{align}
  E_\nu^{\mathrm{min}} = \frac{1}{2} \left(E_r+\sqrt{E_r^2 + 2E_r m_T}\right) \,.
\end{align}
For $m_T \gg E_r$, the case we will mostly be concerned with in this paper, we
can make the approximation $E_\nu^{\mathrm{min}} \simeq \sqrt{m_T E_r/2}$.  The
event rate at a detector is obtained by folding $d\sigma/dE_r$ with the solar
neutrino flux $d\Phi/dE_\nu$:
\begin{align}
  \frac{d R}{dE_r} = N_T \int^\infty_{E_\nu^{\mathrm{min}}}
                     \frac{d\Phi}{d E_\nu} \frac{d\sigma}{dE_r}\, dE_\nu \,,
  \label{eq:event-rate}
\end{align}
where $N_T$ is the number of target particles in the experiment.

In order for a nucleus of mass $m_N$ to receive a recoil energy $E_r
\sim$~few~keV (above the threshold of dark matter detectors), the neutrino must
have an energy $E_\nu \gtrsim \sqrt{E_r m_N / 2} \sim$~few~MeV. Only the high
energy tail of the solar neutrino spectrum exceeds this threshold, so the rate
of nuclear recoil events is low.  On the other hand, the requirement for an
electron recoil above threshold is only $E_\nu \gtrsim \text{few} \times
10$~keV, allowing for most of the solar neutrino flux to contribute. As a
result, the spectrum of solar neutrino-induced electron recoil events in a dark
matter detector with a threshold of few to 10~keV is dominated by the
low-energy $pp$ neutrinos ($E_\nu < 420$~keV), whereas nuclear recoils are
mostly induced by ${}^8$B neutrinos ($E_\nu \lesssim 15$~MeV).

\begin{figure}
  \begin{tabular}{cc}
    \includegraphics[width=8cm]{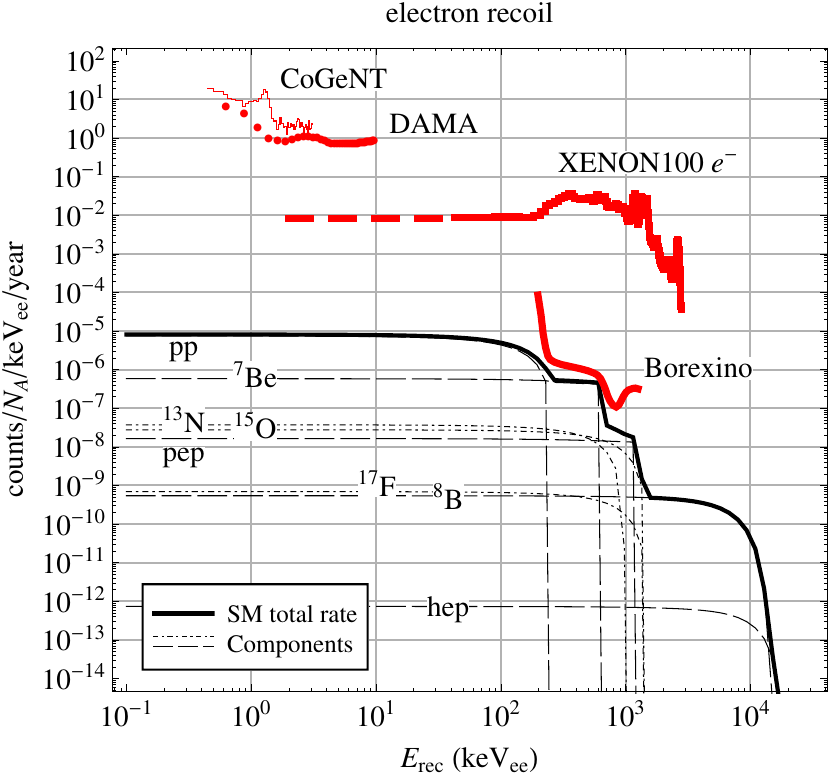} &
    \includegraphics[width=8cm]{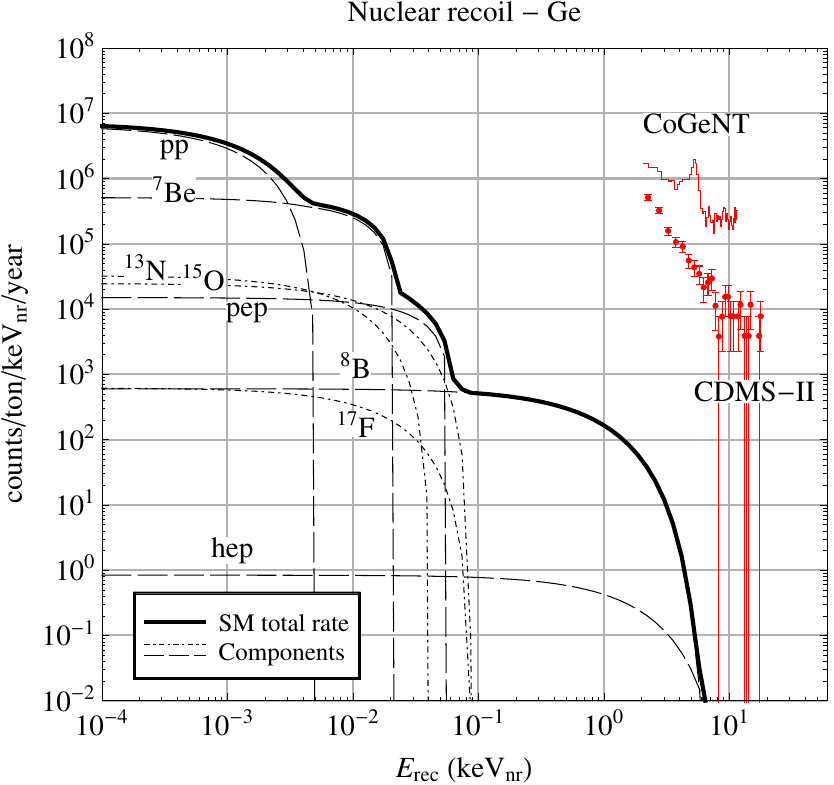} \\
    (a) & (b)
  \end{tabular}
  \caption{Expected event rates in dark matter detectors from (a) solar
    neutrino--electron scattering and (b) solar neutrino--nucleus scattering in
    germanium. In (a), we use units of events per keVee per year per $N_A$ electrons
    (where $N_A$ is the Avogadro number) to be able to compare rates in different
    materials. Thick black lines correspond to the total event rate, while
    thin lines break the rate up into contributions from different neutrino
    production processes.  We also show the observed electron recoil spectra in
    XENON-100~\cite{Aprile:2011vb} (see text for details) and
    Borexino~\cite{Borexino:2011rx}, from the low-threshold analysis of CDMS
    data~\cite{Ahmed:2010wy}, and the event spectra from
    CoGeNT~\cite{Aalseth:2011wp} and DAMA~\cite{Bernabei:2008yi}.  Since CoGeNT
    and DAMA cannot distinguish nuclear recoils from electron recoils, we
    interpret their data as electron recoil in the left panel and as nuclear recoils
    in the right panels.}
  \label{fig:SMspectrum}
\end{figure}

The Standard Model neutrino--electron and neutrino--nucleus scattering rates
are plotted in figure~\ref{fig:SMspectrum} and compared to the event rates
observed in several dark matter detectors and neutrino experiments. Note that the
neutrino--nucleus scattering rate at any given recoil energy depends on the
target material. In figure~\ref{fig:SMspectrum}b, we have chosen germanium as
an example, see~\cite{Gutlein:2010tq} for plots of scattering rates on
different nuclear targets. Note also that we neglect the fact that the electrons and
nuclei in the target material are in bound states.  This is justified because
their binding energies are in most cases much smaller than the
$\mathcal{O}(\text{keV})$ recoil energies we are interested in.\footnote{This is
different for the scattering of dark matter with a mass $\gtrsim 1$~GeV on
electrons, where observable recoil energies can occur only when the electron
enters the scattering process with a large initial momentum of at least a few
MeV. In this case, a dark matter detector would be probing the high-momentum
tail of the bound state electron wave functions, and, consequently, the fact
that electrons are bound cannot be neglected~\cite{Kopp:2009et}.} The only
exception are the very inner electrons in the case of a heavy target material,
which can have binding energies of order 10~keV. We have checked that including
the effects of electron wave functions in our calculation leads only to a
correction to the predicted neutrino--electron scattering rate of at most
few--20\% and introduces small spectral features at those energies where
additional electron shells become kinematically accessible. By neglecting these
small corrections, our results for neutrino--electron scattering become
material-independent and can thus be directly applied to any dark matter direct
detection experiment.

In this paper we will consider a variety of new physics scenarios which add a
new contribution to the differential cross section in the integrand of
equation~\eqref{eq:event-rate}. Obviously we will need to take care and respect all
limits from existing measurements of the solar neutrino flux, as well as other
neutrino experiments. It is thus useful to review the most important
constraints:\\

\noindent\emph{Neutrino--electron interactions:}\\
For neutrino--electron scattering (figure~\ref{fig:SMspectrum}a), the strongest
limits on any anomalous contribution to the cross section comes from
Borexino~\cite{Borexino:2011rx} and SNO~\cite{Aharmim:2005gt, Aharmim:2009gd},
which have energy thresholds of $\sim 250$ keV and 6~MeV, respectively. This
leaves little room for new physics at $E_r \gtrsim 200$~keV. At lower energies
the electron recoil event rates observed in the dark matter experiments
XENON-100~\cite{Aprile:2011vb}, DAMA~\cite{Bernabei:2008yi} and
CoGeNT~\cite{Aalseth:2011wp} lie 3 to 5 orders of magnitude above the Standard
Model solar neutrino background, so that a substantial enhancement of the
neutrino--electron scattering rate cannot be excluded there.  The most
constraining measurement is by XENON-100, but we should stress that the precise
energy threshold and rate for electron recoils below $E_r \lesssim \text{few}
\times 10$~keV in XENON-100 are not precisely known since electron recoils are
not relevant to XENON-100's dark matter search except as a background. The
energy calibration for electron recoils depend on the light yield $L_y$, which
gives the number of detected scintillation photons as a function of the
electron recoil energy, and which was only measured at higher energies
(122~keV) and has to be extrapolated down to lower energies.  In
figure~\ref{fig:SMspectrum}a we have used XENON-100's working assumption that
the light yield at low energies is the same as in the calibration measurement
at $E_r = 122$~keV, $L_y=2.2$~PE/keVee, which leads to a detection threshold
for electron recoils of about 2~keVee.  Measurements indicate that the light
yield might actually be larger at lower recoil energies (down to
30~keVee)~\cite{Sorensen:2008}, and if this trend continues to even lower
recoil energies, the energy threshold in XENON-100 might be even lower (and the
background rate per keVee somewhat higher) than what is shown in
figure~\ref{fig:SMspectrum}a. However, in many scintillators the light yield
peaks at $E_r \sim 10$~keVee and drops steeply below~\cite{Sorensen:2008}, so
that the exact sensitivity of the XENON-100 detector to low-energy electron
recoil events remains somewhat uncertain. In figure~\ref{fig:SMspectrum}, as
well as figures~\ref{fig:Zprime} and~\ref{fig:Zprime-heavy-sterile}, we
indicate this uncertainty by a dashed red lines below $E_r = 50$~keVee.
Besides Borexino and XENON-100, also the GEMMA experiment~\cite{Beda:2009kx}
has placed limits on neutrino--electron scattering at low recoil energies.
GEMMA limits are not directly comparable to to the limits shown in
figure~\ref{fig:SMspectrum} because GEMMA used not solar neutrinos, but reactor
anti-neutrinos, and thus the neutrino spectrum was different. We will comment
more on GEMMA in section~\ref{sec:electron-recoil} when discussing neutrino
magnetic moments, and also in section~\ref{sec:paramspace}.
\\

\noindent\emph{Neutrino--nucleus interactions:}\\
For neutrino--nucleus scattering (figure~\ref{fig:SMspectrum}b), we compare the
Standard Model prediction to the observed event rates in
CoGeNT~\cite{Aalseth:2011wp} and CDMS~\cite{Ahmed:2010wy}.  Here we discuss
only \emph{elastic} neutrino--nucleus scattering because it has been shown
in~\cite{Pospelov:2011ha} that the new physics contributions to the cross
sections for inelastic processes like neutrino-induced deuteron breakup or
nuclear excitations are about eight orders of magnitude smaller than the
elastic scattering cross section in the class of models we are interested in
this paper. The reason is that in our models, deuteron dissociation can only be
mediated by an isoscalar vector current, whereas in the Standard Model, the
isovector axial vector gives by far the dominant contribution.  Thus,
considering  the 4\% uncertainty of the measurement~\cite{Aharmim:2009gd} and
the 16\% uncertainty in the prediction based on the standard solar
model~\cite{Bahcall:2004pz}, no deviations from the Standard Model are expected
in SNO. Note, however, that a dedicated Borexino search for gamma ray lines
from the process $\iso{C}{12} + \nu \to \nu + \iso{C^*}{12} \to \nu_b +
\iso{C}{12} + \gamma$ could be sensitive to the types of new physics discussed
here~\cite{Pospelov:2011ha}.  Also, at much higher energies $\gtrsim 100$~MeV,
there should be constraints from accelerator neutrino experiments such as
MiniBooNE or MINOS. We see from figure~\ref{fig:SMspectrum}b that in germanium
solar neutrinos can only yield neutrino--nucleus scattering events with $E_r
\lesssim 10$~keV because of the sharp drop-off of their spectrum at high
energies. In materials containing lighter target nuclei, the upper end up the
recoil spectrum can be higher, up to 20~keV for NaI and 30~keV for
CaWO$_4$~\cite{Gutlein:2010tq}.  Also, atmospheric neutrinos can induce
scattering events with higher recoil energies~\cite{Gutlein:2010tq}, but
because of their much lower flux, we ignore these events in this paper.

In the following sections, we will discuss how new contributions to
neutrino--electron scattering and new contributions to neutrino--nucleus
scattering can modify the event rates plotted in figure~\ref{fig:SMspectrum}.

%==============================================================================
\section{Four models}
\label{sec:model-intro}
%==============================================================================

We will now turn to new physics scenarios in which neutrino interactions with
electrons and/or nuclei are enhanced to give event rates that are interesting
for present and future dark matter direct detection experiments.  As we have
seen in the previous section, the precise measurements of the solar
neutrino--electron scattering rate above few hundred keV in
Borexino~\cite{Arpesella:2007xf, Arpesella:2008mt, Borexino:2011rx} and
SNO~\cite{Aharmim:2005gt,Aharmim:2009gd} set tight constraints on anomalous
neutrino--electron interactions. Neutrino--nucleus interactions, on the other
hand, should be constrained at energies $\gtrsim 100$~MeV by accelerator
neutrino experiments. Thus, any new physics model that could potentially
contribute significantly to the event rate in direct detection experiments
should give a signal only at low recoil energies around a few keV, but die away
at higher recoil energies.

The simplest possibility which fits the bill is a new interaction between
neutrinos and electrons mediated by a very light or massless particle. In this
case the matrix element for neutrino--electron or neutrino--nucleus scattering
via the light particle contains a factor $q^{-2} = (2 m_T E_r)^{-1}$, where $q$
is the 4-momentum exchange, $E_r$ is the electron recoil energy, and $m_T$ is the
mass of the target electron or nucleus.  This will lead to a differential cross
section which is a falling function of $E_r$.  The new physics contribution can
thus dominate the Standard Model rates at low energies relevant for direct
detection, but not at high energies where neutrino detectors are sensitive. 

%------------------------------------------------------------------------------
\subsection{Neutrino magnetic moments}
\label{sec:mm}
%------------------------------------------------------------------------------

Perhaps the phenomenologically simplest type of new physics leading to enhanced
solar neutrino--electron scattering at low energies through the exchange of a
light particle is a neutrino magnetic moment interaction of the form
\begin{align}
  \mathcal{L}_{\mu_\nu} \supset \mu_\nu \, \bar\nu
    \sigma^{\alpha\beta} \partial_\beta A_\alpha \nu \,,
  \label{eq:L-mm}
\end{align}
where $\mu_\nu$ is the neutrino magnetic moment and $\sigma^{\alpha\beta}$ is
as usual defined in terms of the Dirac matrices as $\sigma^{\alpha\beta} =
\frac{i}{2} [\gamma^\alpha, \gamma^\beta]$.  In this case the light particle
which mediates new interaction is the photon ($A_\alpha$) itself.
We included only dipole--charge interactions in equation~\eqref{eq:L-mm},
but we have checked that dipole--dipole interactions are
negligible compared to dipole--charge interactions at the recoil energies
relevant to dark matter detectors, whereas at higher recoil energies,
Standard Model weak interactions are dominant for all allowed values of
$\mu_\nu$.

While the Standard Model prediction for the loop-induced neutrino magnetic
moment, $\mu_\nu = 3.2 \times 10^{-19} \mu_B \times (m_\nu /
\text{eV})$~\cite{Marciano:1977wx, Giunti:2008ve} (with the Bohr magneton
$\mu_B = \sqrt{4 \pi \alpha} / 2 m_e$, $\alpha$ being the fine structure
constant) is beyond the reach of current and near future experiments, some
extensions of the Standard Model predict sizeable $\mu_\nu$~\cite{Kim:1976gk,
Marciano:1977wx, Beg:1977xz, Georgi:1990se, Czakon:1998rf, Mohapatra:2004ce},
potentially close to the current 90\% CL upper limit $\mu_\nu < 0.32 \times
10^{-10} \mu_B$ from solar and reactor neutrino
experiments~\cite{Nakamura:2010zzi}, in particular
GEMMA~\cite{Beda:2009kx}.\footnote{Published astrophysical constraints can be
up to an order of magnitude stronger~\cite{Beacom:1999wx, Joshipura:2002bp,
Grimus:2002vb, Nakamura:2010zzi, Raffelt:1999gv,
Kuznetsov:2009zm}, but it is difficult to assess the systematic uncertainties
associated with these limits and to assign a confidence level to them.} The
differential neutrino--electron scattering rate through magnetic moment
interactions is given by~\cite{Vogel:1989iv}
\begin{align}
  \frac{d\sigma_{\mu}(\nu e \to \nu e)}{dE_r} = \mu_\nu^2 \alpha
    \bigg(\frac{1}{E_r} - \frac{1}{E_\nu}\bigg) \,,
  \label{eq:sigma-mm}
\end{align}
and the corresponding expression for neutrino--nucleus scattering is
\begin{align}
  \frac{d\sigma_{\mu}(\nu N \to \nu N)}{dE_r} = \mu_\nu^2 \alpha Z^2 F^2(E_r)
    \bigg(\frac{1}{E_r} - \frac{1}{E_\nu}\bigg) \,.
  \label{eq:sigma-mm-N}
\end{align}
Here, $Z$ is the nuclear charge, and $F(E_r)$ is the nuclear form factor (see
discussion below equation~\eqref{eq:dsigmadE-nu-N}).  Of course, ordinary
scattering through $W$ and $Z$ exchange, with the cross section from
equations~\eqref{eq:dsigmadE-nue-e}--\eqref{eq:dsigmadE-nu-N} is also present.
The dependence of equation~\eqref{eq:sigma-mm} on the neutrino energy $E_\nu$
and the recoil energy $E_r$ arises from the interplay of the photon propagator
and the derivative in the magnetic moment interaction vertex,
equation~\eqref{eq:L-mm}.

%------------------------------------------------------------------------------
\subsection{Gauged $B-L$}
\label{sec:B-L}
%------------------------------------------------------------------------------

As we have seen in section~\ref{sec:mm}, a magnetic moment contribution to the
neutrino--electron and neutrino--nucleus scattering cross section falls
proportional to $E_r^{-1}$ at low recoil energy.  We will now turn our
attention to scattering processes for which the recoil energy spectrum falls
even more steeply ($\propto E_r^{-2}$), and hence a larger enhancement of the
neutrino scattering rate at low energies is possible without violating the
Borexino constraint.

This can be achieved if there is a new neutrino--electron or neutrino--quark
interaction mediated by a light particle whose couplings do not contain
derivatives.  Let us in particular consider a model with gauged $B-L$ (baryon
number minus lepton number) symmetry, with the corresponding $U(1)_{B-L}$ gauge
boson $A'$ having a mass $M_{A'} \ll 1$~GeV:
\begin{align}
  \mathcal{L}_{B-L} \supset -g_{B-L} \, \bar{e} \gamma^\alpha A'_{\alpha} e
  + \frac{1}{3} g_{B-L} \, \bar{q} \gamma^\alpha A'_{\alpha} q
    - g_{B-L} \, \bar\nu \gamma^\alpha A'_{\alpha} \nu + \dots \,.
  \label{eq:L-A'}
\end{align}
Here, $g_{B-L}$ is the $U(1)_{B-L}$ coupling constant and $q$, $e$ and $\nu$
are quark, charged lepton and neutrino fields, respectively. We will call $A'$
a ``dark photon'' here and in the following.\footnote{In the literature, the
term ``dark photon'' is often reserved for $U(1)'$ gauge bosons coupling to the
Standard Model only through kinetic mixing with the photon, but we will use it
in a more general context.  In fact, a gauge boson coupled to Standard Model
particles only through kinetic mixing with the photon, would not have tree
level couplings to Standard Model neutrinos at all since they are electrically
neutral.} Note that we neglect the possibility of kinetic mixing between the
dark photon and the photon here. We will discuss models with kinetic mixing
(but with couplings to $B-L$) in great detail below, and we will argue in
section~\ref{sec:paramspace} that, in many phenomenologically relevant processes,
a coupling to $B-L$ is equivalent to kinetic mixing.

The cross section for $A'$-mediated elastic scattering of a neutrino
off an electron or nucleus depends mildly on the chiral structure of
the $A'$ couplings.  Here, for concreteness we will assume the $A'$ to
have pure vector couplings of the form $\bar\psi \gamma^\mu \psi_e \,
A'_\mu$, but spectra would look similar for other chiral
structures. With this assumption we obtain for the differential cross
section for neutrino--electron scattering as a function of the recoil
energy\footnote{We neglected the interference term between $A^\prime$,
  $W$ and $Z$ exchange. As correctly pointed out in
  Ref.~\cite{Bilmis:2015lja}, the bounds derived in this work would be
  improved by about 30\% if the interference terms were taken into
  account.}
\begin{align}
  \frac{d\sigma_{A'}(\nu e \to \nu e)}{dE_r} =
      \frac{g_{B-L}^4 m_e}{4 \pi p_\nu^2 (M_{A'}^2 + 2 E_r m_e)^2}
      \big[ 2 E_\nu^2 + E_r^2 - 2 E_r E_\nu - E_r m_e - m_\nu^2 \big] \,,
  \label{eq:Aprime-xsec}
\end{align}
The corresponding expression for neutrino--nucleus scattering is
straightforwardly obtained by replacing $m_e$ with the nuclear mass, and by
including a coherence factor $A^2$ (where $A$ is the nuclear mass
number) and the nuclear form factor $F^2(E_r)$.\footnote{For recoil energies
below 10~keV, the inclusion of the form factor merely changes the count rate in
a dark matter detector by 1--10\%} Note that in equation~\eqref{eq:Aprime-xsec}
we do not neglect the neutrino mass $m_\nu$ and we distinguish between the
neutrino energy $E_\nu$ and its momentum $p_\nu$ because later, in
section~\ref{sec:electron-recoil}, we will consider also scattering of heavy
sterile neutrinos.

As we will expose in section~\ref{sec:paramspace}, the parameter space for
light $U(1)_{B-L}$ gauge bosons is strongly constrained by ``fifth force''
searches~\cite{Bordag:2001qi, Bordag:2009, Adelberger:2006dh,
Adelberger:2009zz} which require a $U(1)_{B-L}$ gauge boson with couplings
relevant to a dark matter detector to be heavier than about 100~eV. In
addition, bounds on anomalous energy losses in stars and
supernovae~\cite{Redondo:2008aa, Jaeckel:2010ni, Dent:2012mx}, as well as GEMMA
limits on anomalous contributions to the scattering of reactor antineutrinos on
electrons~\cite{Beda:2009kx} severely limit the allowed range of coupling
constants for $1\ \text{meV} \lesssim M_{A'} \lesssim 100\ \text{MeV}$.
However, we will see that some interesting regions of parameter space are not
yet fully excluded. Moreover, in slightly non-minimal models, many constraints
can be easily avoided~\cite{Feldman:2006wg, Nelson:2007yq}.

%------------------------------------------------------------------------------
\subsection{Sterile neutrinos and a dark photon coupled through kinetic mixing}
\label{sec:kinetic-mixing}
%------------------------------------------------------------------------------

One simple possibility to avoid many constraints on light new gauge bosons
while still maintaining large neutrino--electron and neutrino--nucleus
scattering rates is to consider scenarios in which the couplings of the new
gauge boson to electrons and other Standard Model particles are much smaller
than its coupling to at least some neutrino flavors. This is possible in models
with new sterile neutrinos, which are singlets under the Standard Model gauge
group, but charged under a new $U(1)'$ gauge group. Standard Model particles,
on the other hand, could be coupled to the $A'$ gauge boson (which we again
call the ``dark photon'') only through a small kinetic mixing $\epsilon$ with
the photon.  The relevant terms in the Lagrangian of the model below the
electroweak scale are
\begin{align}
  \mathcal{L} &\supset
    -\frac{1}{4} F'_{\mu\nu} F'^{\mu\nu} - \frac{1}{4} F_{\mu\nu} F^{\mu\nu}
           - \frac{1}{2} \eps F'_{\mu\nu} F^{\mu\nu}
           + \bar{\nu}_{s} i\slashed{\partial} \nu_{s}
           + g' \bar{\nu}_{s} \gamma^\mu \nu_{s} A'_\mu  \nonumber\\
    &\quad - \overline{(\nu_L)^c} m_{\nu_L} \nu_L 
           - \overline{(\nu_s)^c} m_{\nu_s} \nu_s 
           - \overline{(\nu_L)^c} m_{\mathrm{mix}} \nu_s \,,
  \label{eq:L-light-steriles}
\end{align}
where the first line contains the gauge kinetic terms of the $A'$ boson and the
photon, the kinetic mixing term between $A'$ and the photon with small mixing
parameter $\eps$, and the kinetic term of the sterile neutrino $\nu_s$,
including its gauge coupling to the $A'$ with coupling constant $g'$. (We have
omitted the kinetic terms of the other fermions and gauge bosons in the model
since they are not relevant to our discussion and are anyway unchanged compared
to the Standard Model.) We could equivalently have considered the interactions
of $A'$ above the electroweak scale, in which case the kinetic mixing would be
between the dark photon and the hypercharge boson rather than the photon. This
would introduce not only mixing between $A'$ and $A$, but also mixing between
$A'$ and the $Z$, which is, however, negligible for the low energy processes we
are interested in. The second line of equation~\eqref{eq:L-light-steriles}
contains the usual $3 \times 3$ Majorana mass matrix for active neutrinos
$\nu_L$, a Majorana mass matrix for the sterile neutrinos, as well as a
Majorana-type mixing term.  These mass matrices may be obtained from a seesaw
mechanism if we introduce singlet right-handed neutrinos $\nu_R$:
\begin{equation}
\label{eq:seesaw}
  - \bar{L} Y_\nu \tilde{H} \nu_R - \bar{\nu}_{s} Y_s H' \nu_R
         - \frac{1}{2} \overline{(\nu_R)^c} M_R \nu_R + h.c. \,,
\end{equation}
where all fields are understood as vectors in flavor space: $\bar{L}$ contains
the three Standard Model lepton doublets, $\nu_s$ contains $n_s$ sterile
neutrinos coupled to $A'$, and $\nu_R$ contains $3 + n_s$ heavy right-handed
neutrinos. We use the notation $H$ for the Standard Model Higgs boson, $H'$ for
a new Higgs boson which is charged under $U(1)'$ and breaks it when it acquires
a vacuum expectation value (vev), and we define $\tilde{H} \equiv \eps^{ab}
H_b^\dag$, where $\eps^{ab}$ is the totally antisymmetric tensor in two
dimensions and $a$, $b$ are $SU(2)_L$ indices.\footnote{We neglect the ``Higgs portal''
coupling $(H^\dag H)(H'^\dag H')$ here.} The Standard Model Yukawa
coupling $Y_\nu$,  the Yukawa coupling of the sterile neutrinos $Y_s$, and the
right-handed mass matrix $M_R$ are understood to be matrices of size $3 \times
(3+n_s)$, $n_s \times (3+n_s)$, and $(3+n_s) \times (3+n_s)$, respectively.
According to the usual seesaw formula, the effective $(3+n_s) \times (3+n_s)$
Majorana mass matrix of the light neutrinos is given by
\begin{align}
  m_\nu = \begin{pmatrix} Y_\nu \ev{H} \\ Y_s \ev{H'} \end{pmatrix}
    M_R^{-1} \begin{pmatrix} Y_\nu \ev{H} \\ Y_s \ev{H'} \end{pmatrix}^T \,.
  \label{eq:m-nu}
\end{align}
We have not explicitly written down the kinetic and potential terms for $H$ and
$H'$, but we assume that they are such that $H'$ acquires a small vev that
gives the $A'$ a mass consistent with the constraints from
section~\ref{sec:paramspace}, and the sterile neutrinos a mass that is
sufficiently small to allow coherent mixing between active and sterile flavors
at typical solar neutrino energies. Since the vev of $H'$, denoted by
$\ev{H'}$, is by assumption much smaller than the Standard Model Higgs vev
$\ev{H}$, the mostly sterile neutrino mass eigenstate will typically be even
lighter than the active ones unless there is a large hierarchy in the
right-handed Majorana mass matrix $M_R$.

In models of the form \eqref{eq:L-light-steriles}, the sterile
neutrino--electron scattering cross section is given by
equation~\eqref{eq:Aprime-xsec}, with the replacement $g_{B-L} \to
\sqrt{\epsilon e g'}$. Here, $\epsilon e$ denotes the $A'$ coupling to
electrons, and $g'$ denotes the $U(1)'$ gauge coupling constant.
(If the dark photon is \emph{very} light, so that the range of its
interaction becomes macroscopic, there will be corrections
to equation~\eqref{eq:Aprime-xsec} at low recoil energy due
to the breakdown of the one-boson-exchange
approximation and due to possible shielding effects from a cosmic
sterile neutrino background~\cite{Dolgov:1995hc}.)
We will see
that in such models, $d\sigma_{A'}(\nu e \to \nu e)/dE_r$ and $d\sigma_{A'}(\nu
N \to \nu N)/dE_r$ can be significantly larger than in models with only the
three active neutrinos.

A small admixture of sterile neutrinos to the solar neutrino flux can be
produced by oscillation before the neutrinos reach the Earth. In a two-flavor
approximation with only one active neutrino flavor $\nu_a$ and one sterile
neutrino flavor $\nu_s$, and with the corresponding mass eigenstates $\nu_2$
and $\nu_4$, the vacuum oscillation probability is given by the usual
expression
\begin{equation}
  P(\nu_a \to \nu_s) = \sin^2 2 \theta_{24}
                       \sin^2\bigg( \frac{\Delta m^2_{42} L}{4E} \bigg) \,,
  \label{eq:osc-prob}
\end{equation}
where $L$ is the distance traveled by the neutrinos, $\theta_{24}$ is the
effective active--sterile neutrino mixing angle in vacuum, and $\Delta m_{42}^2
= m_4^2 - m_2^2$ is the splitting between the squared mass of the mostly
sterile mass eigenstate ($m_4$) and the mostly active mass eigenstate ($m_2$)
in vacuum.  The fraction of sterile flavors in the solar neutrino flux can be
as large as 20--30\% without violating SNO constraints on the rate of neutral
current neutrino--nucleon interactions~\cite{Aharmim:2011yq}. If the mass
squared difference between active and sterile neutrinos is in the range $\Delta
m_{41}^2 \gtrsim 10^{-4}$~eV$^2$ accessible to terrestrial neutrino oscillation
experiments, some of the active--sterile mixing angles are more strongly
constrained, to the level of $\sin^2 2\theta \sim \text{few
\%}$~\cite{Kopp:2011qd, Giunti:2011gz, Giunti:2011hn}.

One potential problem with the model discussed here is that the sterile
neutrinos (and also the hidden Higgs boson $H'$~\cite{Ahlers:2008qc}) will
acquire a tiny effective electromagnetic charge of order
$\epsilon$~\cite{Holdom:1985ag}. \footnote{This is true in a basis where the
kinetic mixing term has been transformed away by the replacement $A'_\mu \to
A'_\mu - \eps A_\mu$ and the mass matrix of the two $U(1)$ gauge bosons is
hence off-diagonal. In parts of the literature, a basis with a diagonal mass
matrix is used, in which case no electromagnetic minicharges occur. Of course,
the constraints discussed here are basis-independent.}
There are tight limits on such ``minicharged''
particles, the strongest of which come from bounds on anomalous energy losses
in stars due to the decay of plasmons into sterile
neutrinos~\cite{Davidson:2000hf,Jaeckel:2010ni}.\footnote{Plasmon oscillations
and plasmon decay into sterile neutrinos is also an alternative production
mechanism for sterile neutrinos in the Sun.  However, the $A'$ spectrum and
hence also the sterile neutrino spectrum in this case are determined by thermal
effects and are steeply falling functions of energy above
1~keV~\cite{Redondo:2008aa}. We have checked that therefore the flux at
energies accessible to dark matter experiments is negligibly small.} Like
bounds on dark photons, these constraints are also somewhat model dependent as
we will discuss in section~\ref{sec:paramspace}. For us, it is important to
note that minicharged particles with sufficiently \emph{large} coupling cannot
leave a stellar environment and may therefore still be allowed. Also, models
which raise the sterile neutrino mass to $\sim 10$~keV to several 100~keV
(depending on $\epsilon$) completely evade these limits.

Note that there are other constraints on sterile neutrinos with masses $\gtrsim
10$~KeV, the most severe of which come from observations of the x-ray flux from
galaxy clusters and of the cosmic microwave background,
see~\cite{Smirnov:2006bu, Kusenko:2009up}. These observables are modified if
there is a large astrophysical population of sterile neutrinos decaying
radiatively into light neutrinos. Other constraints could be derived from
requiring the sterile neutrinos to not overclose the Universe.  However,
astrophysical and cosmological bounds can be avoided if sterile neutrinos are
not produced in significant numbers in the early universe, for instance because
of a low reheating temperature ($\ll 100$~MeV)~\cite{Gelmini:2004ah}, or
because they are chameleon-like (i.e.\ their effective mass could depend on the
surrounding matter density~\cite{Nelson:2007yq, Nelson:2008tn,
Feldman:2006wg}).  Alternatively, the sterile neutrinos could have a fast
invisible decay mode, for instance into light neutrinos and
Majorons~\cite{Smirnov:2006bu}.

The phenomenology of $\gtrsim 10$~keV sterile neutrinos in a direct detection
experiment is essentially the same as that of the light sterile neutrinos
discussed above, but their production in the Sun can be modified, especially
due to kinematic suppression and because they can no longer interfere with the
light active neutrinos.  We discuss the construction of models with heavier
sterile neutrinos in more detail in appendix~\ref{sec:heavy-nus-model}, and
their interesting phenomenology in section~\ref{sec:e-recoil-heavy}.

%------------------------------------------------------------------------------
\subsection{Baryonic sterile neutrinos and gauged baryon number}
\label{sec:gauged-B}
%------------------------------------------------------------------------------

While the models discussed so far are most easily detected in
neutrino--electron scattering, it is phenomenologically interesting to
consider also scenarios which predict neutrino--nucleus scattering to
be dominant, in particular since nuclear recoils are the type of
signal that most dark matter detectors are particularly sensitive
to. A model of this type has been proposed in~\cite{Pospelov:2011ha}:
It extends the Standard Model by a gauged $U(1)_B$ (baryon number)
symmetry and introduces one or several ``baryonic'' sterile neutrinos
which are charged under $U(1)_B$. Since constraints on new light gauge
bosons coupling to quarks are much weaker than constraints on
particles coupling to leptons (see section~\ref{sec:paramspace} below),
the $U(1)_B$ gauge coupling can be fairly large.

The cross section for sterile neutrino--nucleus scattering in this model is
given by equation~\eqref{eq:Aprime-xsec}, with $g_{B-L}$ replaced by the
$U(1)_B$ gauge coupling $g_B$, and with the usual coherence factor
$A^2$ and the nuclear form factor $F^2(E_r)$ included.

In addition to enhancing neutrino scattering cross sections, a $U(1)_B$ gauge
boson is also a source of new Mikheyev-Smirnov-Wolfenstein (MSW) type neutrino
matter effects~\cite{Mikheyev:1986wj, Wolfenstein:1977ue}. In particular,
$A'$-mediated coherent forward scattering of sterile neutrinos on nucleons
creates a potential of the form
\begin{align}
\label{eq:matterpotential}
  V_{A'} = \frac{g_B^2}{M_{A'}^2} \, (N_p + N_n) \,
\end{align}
where $N_p$ and $N_n$ are the proton and neutron number densities in the Sun,
respectively, which are a function of the distance from the
center~\cite{Bahcall:2004pz}. Note that this form of the MSW potential is
specific to the $U(1)_B$ model. The $U(1)'$ model discussed in
section~\ref{sec:kinetic-mixing}, for instance, does \emph{not} lead to
non-standard matter effects at all because the couplings of a kinetically mixed
$U(1)'$ gauge boson are proportional to electric charge, which is zero for
ordinary matter, whereas a $U(1)_{B-L}$ model leads to a potential proportional
to the number density of neutrons. (In the $U(1)_{B-L}$ model from
section~\ref{sec:B-L}, this matter potential is however inconsequential for
oscillation physics since it is flavor-diagonal among the active neutrinos.)

The phenomenological implications of the new MSW potential are twofold: On the
one hand, in the large potential limit
\begin{align}
  V_{A'} \gg \max_{j,k} |\Delta m_{jk}^2| / 2E \,, \qquad\qquad
  \text{$j$, $k = 1 \dots$number of neutrino mass eigenstates} \,,
  \label{eq:V-condition}
\end{align}
the Hamiltonian terms mixing active and sterile neutrinos are negligible
compared to the large potential term for the sterile neutrinos.  (Here, $E$ is
the neutrino energy, which is $\gtrsim 10$~MeV for the neutrinos of interest to
dark matter detectors.) Thus, in this case, the effective active--sterile
mixing angles in matter are very small, and sterile neutrinos are effectively
decoupled in matter. Their production in the Sun is then negligible, and they
are only produced via vacuum oscillations outside the Sun.

On the other hand, even when active--sterile oscillations are suppressed, the
existence of the $V_{A'}$ term can still modify oscillations among the three
active neutrino flavors. For each parameter point considered, one needs to check
that the agreement of the theory with data from solar, atmospheric, reactor,
and accelerator neutrino experiments is not spoiled.  For solar neutrinos, this
requires checking that the $\nu_e \to \nu_e$ survival probability between $\sim
200$~keV--$20$~MeV is not modified substantially compared to the standard
three-flavor case. Since in some models, Earth matter effects can lead to
different survival probabilities during daytime and nighttime, one should also
make sure that no such day--night asymmetry is present. For terrestrial
neutrinos, one has to check that the $\parenbar{\nu}_e \to \parenbar{\nu}_e$
and $\parenbar{\nu}_\mu \to \parenbar{\nu}_\mu$ survival probabilities, as well
as the $\parenbar{\nu}_\mu \leftrightarrow \parenbar{\nu}_e$ and
$\parenbar{\nu}_\mu \to \parenbar{\nu}_\tau$ oscillation probabilities are not
modified substantially.

Finally, the existence of a non-standard matter potential could imply the
existence of new MSW resonances which may lead to strong conversion of active
neutrinos into sterile states. These MSW resonances can usually be avoided by
choosing an appropriate sign for the active--sterile mass differences.

%==============================================================================
\section{Enhanced neutrino--electron scattering from new physics}
\label{sec:electron-recoil}  
%==============================================================================

Let us now investigate the phenomenology of the models introduced in
section~\ref{sec:model-intro} in more detail. We begin by studying
neutrino--electron scattering rates in dark matter detectors. Most of these
experiments make an effort to distinguish nuclear and electron recoils,
focusing on the former as dark matter candidate events and rejecting the latter
as backgrounds.  Interestingly, two exceptions to this are
DAMA~\cite{Bernabei:2008yi} and CoGeNT~\cite{Aalseth:2010vx}, both of which
have observed a possible signal.  We will begin with scenarios in which the
scattering neutrinos (either active or sterile) are light ($\lesssim 1$~eV),
and later consider also heavier sterile neutrinos.

%------------------------------------------------------------------------------
\subsection{Scattering of Light Neutrinos}
\label{sec:e-recoil-light}
%------------------------------------------------------------------------------

\begin{figure}
  \begin{center}
    \includegraphics[width=8cm]{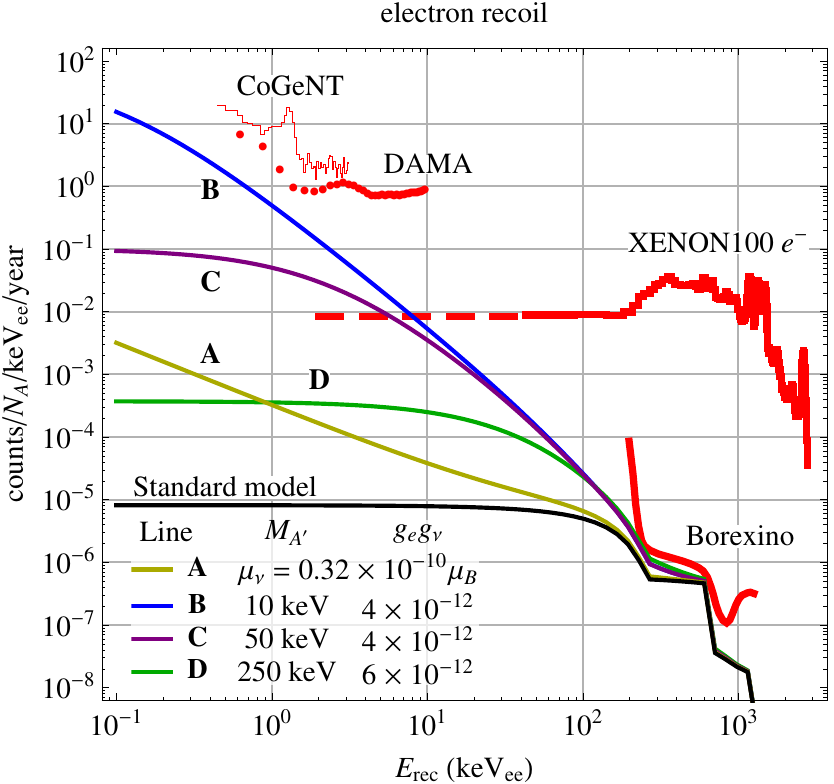} 
  \end{center}
  \caption{Expected event spectra in a dark matter detector from new physics in
    the scattering of solar neutrinos on electrons.  The different colored
    curves correspond to (A) a model where the neutrino has a magnetic dipole
    moment of $\mu_\nu=0.32\times 10^{-10} \mu_B$ and (B, C, D) models where
    the scattering is enhanced by the exchange of a new light gauge boson $A'$
    with couplings $g_e$ to electrons and $g_\nu$ to neutrinos. The latter case
    is for instance realized in the model from
    section~\ref{sec:kinetic-mixing}, where Standard Model particles couple to
    the $A'$ through its kinetic mixing with the photon, but there is also a
    sterile neutrino $\nu_s$ directly charged under $U(1)'$.  To keep the
    discussion general, we assume the $\nu_e \to \nu_s$ transition probability
    to be energy-independent, and we have absorbed the corresponding flux
    suppression into a redefinition of $g_\nu^2$.  The black curve shows the
    Standard Model rate from figure~\ref{fig:SMspectrum}, and the red curves
    and data points show the observed electron recoil rates in
    XENON-100~\cite{Aprile:2011vb} (see section~\ref{sec:rates} for details),
    Borexino~\cite{Borexino:2011rx}, CoGeNT~\cite{Aalseth:2011wp}, and
    DAMA~\cite{Bernabei:2008yi}. (Note that CoGeNT and DAMA cannot distinguish
    nuclear recoils from electron recoils, so their data can be interpreted as
    either.)}
  \label{fig:Zprime}
\end{figure}

Curve~A in figure~\ref{fig:Zprime} shows the neutrino--electron scattering rate
expected for neutrinos with a magnetic moment (section~\ref{sec:mm}) of $0.32
\times 10^{-10} \mu_B$, saturating the 90\% C.L.\ limit from the GEMMA
experiment~\cite{Beda:2009kx}. We see that a significant enhancement of the
event rate, by more than one order of magnitude at $E_r \sim$~few~keV, is
possible. While this is still outside the reach of existing experiments, near
future detectors like LUX, XENON-1T, X-MASS or PANDA-X may be able to enter
this territory because the self-shielding capabilities of large liquid noble
gas detectors are expected to lead to a significant reduction in radioactive
background levels. Once the uncertainty on the background rate drops below the
expected signal rate, future dark matter detectors may be able to improve the
bounds on the magnetic dipole moment of the neutrino considerably.

Curves~B, C and D in figure~\ref{fig:Zprime} are typical event spectra from
$A'$-mediated neutrino--electron scattering in a dark matter detector.  We see
that, as expected, the electron recoil energy spectrum is proportional to the
squared propagator of the light gauge boson, $(q^2-M_{A'}^2)^{-2}$ where $q^2 =
-2 E_r m_e$.  It is thus a steeply falling function of $E_r$ for $E_r >
M_{A'}^2/ 2 m_e$ and flattens out for lower $E_r$. This can be easily discerned by
comparing curves B, C and D, which where computed assuming different values for
$M_{A'}$. All three of these curves satisfy (but almost saturate) the Borexino
limit, and all of them may be within the reach of LUX, XENON-1T, X-MASS, PANDA-X or even
XENON-100, provided the detector response to electron recoils can be
sufficiently well understood, and the electron recoil background from Standard
Model processes can be sufficiently reduced.  Even now, XENON-100 disfavors
scenarios in which \emph{all} of the events seen in CoGeNT or DAMA are
explained by the scattering of solar neutrinos on electrons.  It is of course
still conceivable that only a fraction of these rates signifies the scattering
of sterile neutrinos on electrons, and the rest is due to instrumental
backgrounds. In fact, a recent preliminary investigation by the CoGeNT
collaboration~\cite{Collar-TAUP} suggests that such instrumental backgrounds
exist. The annual modulation amplitudes observed by DAMA and CoGeNT are roughly
at the level of the XENON-100 background, and we may hope to explain these
signals in models that predict a strongly modulating signal, see
section~\ref{sec:modulation}.

Curves~B, C and D in figure~\ref{fig:Zprime} were computed with the
$U(1)_{B-L}$ model from section~\ref{sec:B-L} and the $U(1)'$ model with
kinetic mixing and $U(1)'$-charged sterile neutrinos from
section~\ref{sec:kinetic-mixing} in mind.  The former model is more strongly
constrained since it predicts enhanced scattering rates of even the active
neutrinos at low energies. Such signals are constrained by low-energy
neutrino--electron scattering experiments at nuclear reactors, in particular
GEMMA~\cite{Beda:2009kx}, and they may also be in conflict with constraints on
any anomalous energy loss in the Sun and in other stars.  (see
section~\ref{sec:paramspace}). Astrophysical constraints can often be avoided
in ``chameleon'' models, where the dark photon mass depends on the background
matter density~\cite{Nelson:2007yq, Nelson:2008tn, Feldman:2006wg}.

The $U(1)'$ model with kinetic mixing allows for much more model-building
freedom than the $U(1)_{B-L}$ model as long as it is ensured that the
oscillation lengths for transitions of active neutrinos into the more strongly
interacting sterile neutrinos,
\begin{align}
  L_{\rm osc} = \frac{4\pi E}{\Delta m_{4i}^2} \simeq
                2.48 \ \text{km} \times \bigg(\frac{E}{\text{MeV}}\bigg)
                       \bigg( \frac{\text{eV}^2}{\Delta m_{4i}^2} \bigg) \,,
  \qquad (i = 1, 2, 3) \,,
\end{align}
are much longer than the distance of the GEMMA detector from the reactor core
(13.9~m~\cite{Beda:2009kx}). Here, we have assumed that this is the case,
but that at the same time
\begin{align}
  L_{\rm osc} \ll 1\ \text{AU}
  \label{eq:Losc-small}
\end{align}
in the neutrino energy range relevant to dark matter detectors, $10\ \text{keV}
\lesssim E_\nu \lesssim 15\ \text{MeV}$, so that the $L$-independent
oscillatory term in equation~\eqref{eq:osc-prob} averages to $1/2$ and the
fraction of solar neutrinos converted into sterile states, $P(\nu_e \to \nu_s)$, is
energy-independent.  Condition~\eqref{eq:Losc-small} is fulfilled for $\Delta
m_{4i}^2 \gg 2.5\times 10^{-10}$~eV$^2$ ($i=1,2,3$). To avoid astrophysical
constraints on kinetically mixed $U(1)'$ gauge bosons and on sterile neutrinos
more easily, one can (but does not have to, see section~\ref{sec:paramspace})
consider sterile neutrinos heavier than $\sim 10$~keV
or chameleon models~\cite{Nelson:2007yq, Nelson:2008tn, Feldman:2006wg}.

Note that in the $U(1)'$ model with kinetic mixing (or in any model where the
signals shown in figure~\ref{fig:Zprime} originate from the scattering of
sterile neutrinos), the product of couplings $g_\nu g_e$ needed to obtain the
displayed curves needs to be larger in order to compensate for the mixing
angle-suppressed flux. This is, however, easily achieved by increasing the
coupling of dark photons to sterile neutrinos while leaving the coupling to
electrons small, or even reducing it compared to the $U(1)_{B-L}$ case.  In the
legend of figure~\ref{fig:Zprime} it is understood that in models where the
enhanced event rate is due to sterile neutrinos, the fraction of solar
neutrinos converted into sterile states, $P(\nu_e \to \nu_s)$, is absorbed into
a redefinition of $g_\nu^2$.

%------------------------------------------------------------------------------
\subsection{Heavy Sterile Neutrinos}
\label{sec:e-recoil-heavy}
%------------------------------------------------------------------------------

In the scenarios shown in figure~\ref{fig:Zprime}, the electron recoil spectra
in the few--10~keV region were falling at most as $E_r^{-2}$, so that a
scattering rate larger than the one shown in curve~B is excluded by Borexino,
and possibly also by XENON-100 constraints. However, we note that the recoil
spectrum has several sharp edges at higher energies of several hundred keV or
above. These features are induced by the kinematic cutoffs of the individual
processes contributing to the solar neutrino flux (see
figure~\ref{fig:SMspectrum} for a break-down of the recoil spectrum into these
contributions).  We will now show how these kinematic edges can be shifted down
to the recoil energies of interest to dark matter detectors. This can happen in
models with heavy sterile neutrinos whose mass is close to one of the kinematic
edges in the solar neutrino flux. Here, we will in particular focus on the
sharpest of these edges, coming from neutrinos produced in the reaction
\begin{align}
  \iso{Be}{7} + e^- \to \iso{Li}{7} + \nu_i \,,
  \label{eq:Be7}
\end{align}
with an energy of 862~keV.

For sterile neutrinos with a mass of that order, the velocity difference
between the heavy and light states produced in the Sun is so large that the
neutrino flux can be viewed as a completely incoherent mixture of heavy and
light states. Hence, dynamic transitions between the active and sterile flavor
eigenstates during propagation are absent. (There can still be coherence among
the light states and among the heavy states, provided the mass splittings
within each of the two sectors are sufficiently small.) We will assume that the
electron recoil signal is dominated by sterile neutrino scattering, as is the
case for instance in the $U(1)'$ model from section~\ref{sec:kinetic-mixing}.
We will also assume that the sterile flavor admixture to the light mass
eigenstates is negligible, whereas there needs to be some admixture of the
heavy mass eigenstate $\nu_4$ to the active flavor eigenstates in order to
produce $\nu_4$ in the Sun. See Appendix~\ref{sec:heavy-nus-model} for a
discussion of models in which this is naturally achieved.

\begin{figure}
  \begin{center}
    \includegraphics[width=8cm]{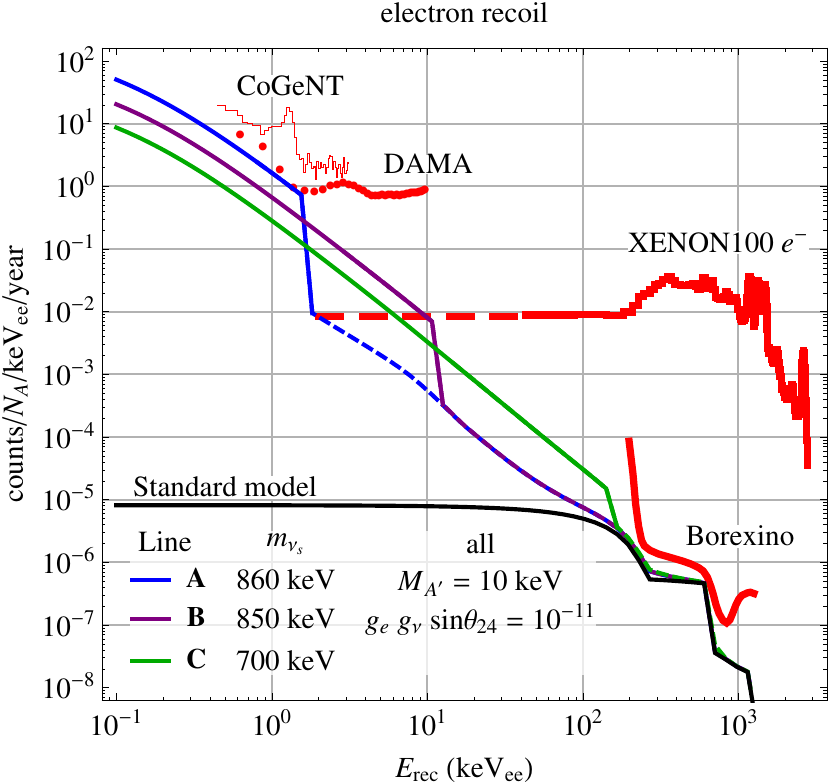}
  \end{center}
  \caption{Expected event spectra in a dark matter detector from
    $A'$-enhanced scattering of heavy sterile neutrinos on electrons (thick
    colored lines, see figure legend for the parameters used). We have assumed
    the $A'$ mass to be almost negligible, and we have chosen the cross section such
    that the CoGeNT excess can be explained.  Black lines show the count rate
    in the Standard Model, and red curves show the observed event rates in
    XENON-100~\cite{Aprile:2011vb} (see section~\ref{sec:rates} for details),
    Borexino~\cite{Borexino:2011rx}, CoGeNT~\cite{Aalseth:2011wp} and
    DAMA~\cite{Bernabei:2008yi}.  We have accounted for the kinematic
    suppression of heavy neutrino production for the $\iso{Be}{7}$ neutrinos
    (solid colored lines), but not of the pp, \iso{N}{13}, \iso{O}{15},
    \iso{F}{17}, \iso{B}{8} and hep neutrinos (dashed colored line), which are
    produced as parts of 3-body final states.}
  \label{fig:Zprime-heavy-sterile}
\end{figure}

Examples for the expected electron recoil rates from scattering of heavy
sterile neutrinos with a mass not far below the \iso{Be}{7} line are shown in
figure~\ref{fig:Zprime-heavy-sterile}.  Even though production of the heavy
mass eigenstate $\nu_4$ in the reaction \eqref{eq:Be7} is suppressed by the
small mixing matrix element $|U_{e4}|^2$ and by kinematic factors which are
small for $\nu_4$ masses close to the $Q$ value of the production reaction, a
considerable enhancement of the neutrino--electron scattering rate in dark
matter experiments is still possible.  Note that in the plot we have not taken
into account kinematic modifications to solar neutrino production processes
other than \eqref{eq:Be7}, many of which have 3-body final states.  Instead, we
have simply assumed the $\nu_4$ produced in these reactions to have the same
energy spectrum (suppressed only by the small leptonic mixing matrix element
$|U_{e4}|^2$) as the corresponding light neutrinos. This introduces
inaccuracies close to the upper kinematic thresholds of the pp, \iso{N}{13},
\iso{O}{15}, \iso{F}{17}, \iso{B}{8} and hep neutrino spectra, and to remind
the reader of these inaccuracies we show the corresponding contributions to
neutrino--electron scattering as dashed lines in
figure~\ref{fig:Zprime-heavy-sterile}.

The plot shows that a model with heavy sterile neutrinos can explain an excess
in low-threshold experiments like CoGeNT, while avoiding constraints from
higher-energy detectors such as XENON-100 and Borexino.  A certain amount of
fine-tuning is, however, required in the particular case of CoGeNT since the
CoGeNT and XENON-100 thresholds are relatively close to each other. Such
tuning of a mass splitting against an unrelated quantity (the $^7$Be line
energy in this case) is reminiscent of tunings of the mass splitting against
the dark matter kinetic energy in models of inelastic dark
matter~\cite{TuckerSmith:2001hy, Chang:2010pr}, which were also introduced to
reduce tensions among direct detection experiments.

The allowed masses and mixing angles of sterile neutrinos are constrained by a
large number of terrestrial and astrophysical experiments, as explained at the
end of section~\ref{sec:kinetic-mixing}, but as also discussed there, these
constraints can be avoided in slightly non-minimal scenarios.

%==============================================================================
\section{Enhanced neutrino--nucleus scattering from new physics}
\label{sec:nuclear-recoil}  
%==============================================================================

While neutrino--electron scattering in a dark matter detector, as discussed in
the previous section, is a very interesting discovery channel for new physics
in the neutrino sector, it is not the process that most of these detectors are
designed to look for. Let us therefore now turn our attention to
neutrino--nucleus scattering, focusing in particular on scenarios in which the
scattering rate at low energies is enhanced, thus possibly mimicking a dark
matter signal.

\begin{figure}
  \begin{center}
    \begin{tabular}{cc}
      \includegraphics[width=8cm]{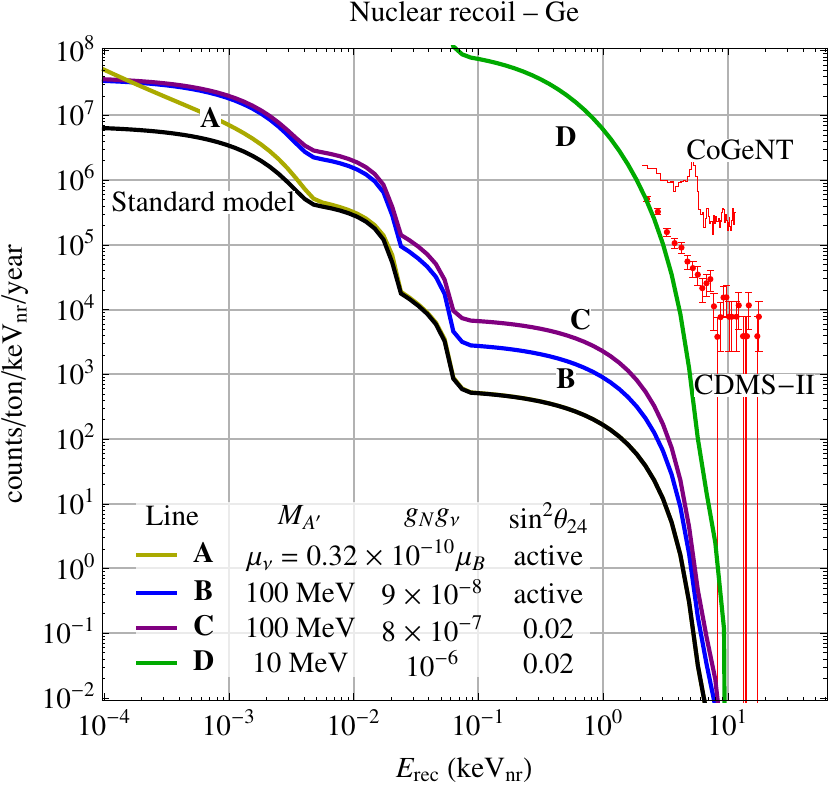} &
      \includegraphics[width=8cm]{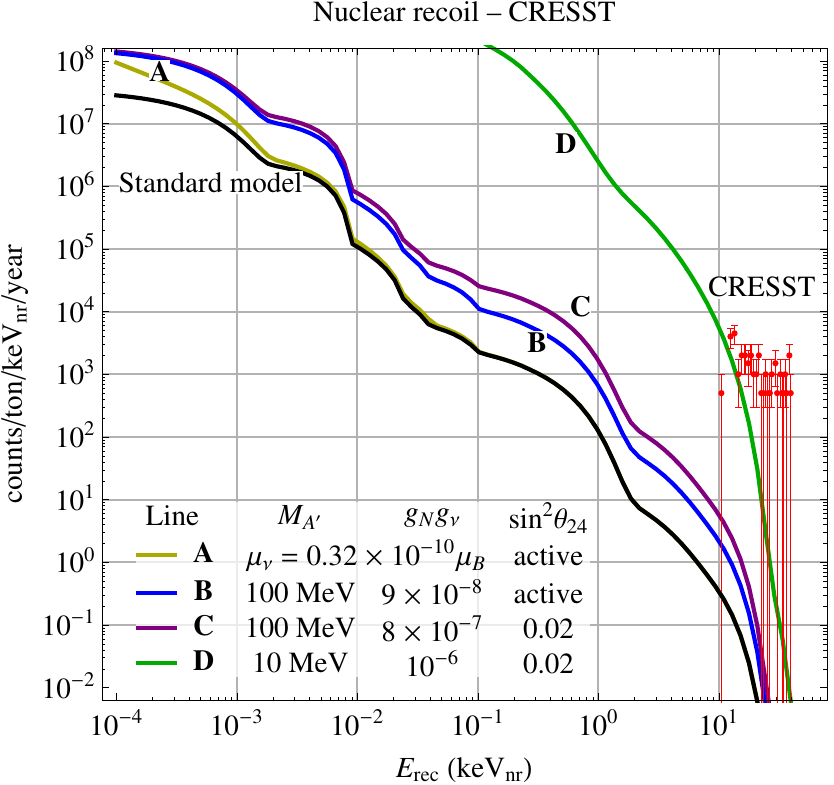} \\
      (a) & (b) \\[0.3cm]
      \includegraphics[width=8cm]{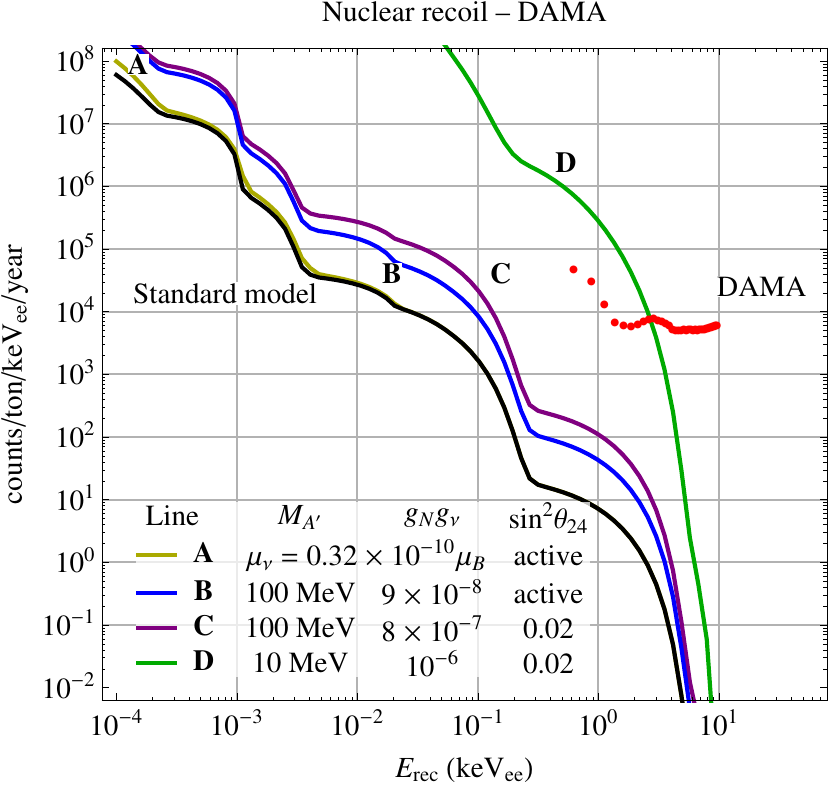} &
      \includegraphics[width=8cm]{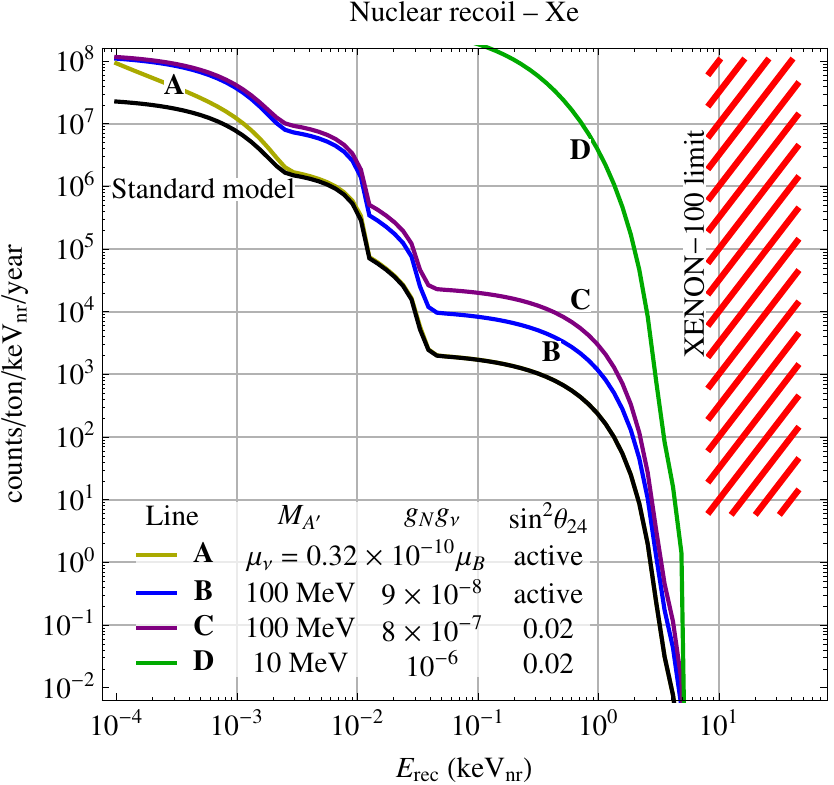} \\
      (c) & (d)
    \end{tabular}
  \end{center}
  \caption{Expected event spectra in dark matter detectors from solar
    neutrino--nucleus scattering in (a) germanium, (b) CaWO$_4$, (c) NaI(Tl),
    and (d) xenon.  Note that for NaI(Tl), we use units of keVee rather than
    keVnr for $E_r$ because due to the different quenching factors for Na (0.3)
    and I (0.09)~\cite{Bernabei:1996vj}, the nuclear recoil energy cannot be
    uniquely reconstructed. Colored curves correspond to (A) a scenario with a
    neutrino magnetic moment $\mu_\nu = 0.32 \times 10^{-10} \mu_B$, (B) a
    model with active neutrino--nucleus scattering through a light $A'$ boson
    (for instance the $U(1)_{B-L}$ model from section~\ref{sec:B-L}), and (C),
    (D) a model in which 2\% of the solar neutrino flux oscillate into a
    Standard Model singlet $\nu_s$, which couples to atomic nuclei for instance
    via a light $U(1)_B$ gauge boson
    (section~\ref{sec:gauged-B})~\cite{Pospelov:2011ha}.  The relevant model
    parameters, in particular the mass of the $A'$ and its coupling to nucleons
    ($g_p = g_n \equiv g_N$) are listed in the legend.  Where applicable, we
    have assumed active--sterile mixing with $\sin^2\theta_{24} = 0.02$ and
    $\Delta m^2_{42} = 10^{-10}$~eV$^2$.  The black curves show the Standard
    Model rate, and the red curves and data points show the observed spectra of
    nuclear recoil candidates in CoGeNT~\cite{Aalseth:2011wp}, in the
    low-threshold data set from CDMS~\cite{Ahmed:2010wy}, in
    CRESST~\cite{Angloher:2011uu}, and in DAMA~\cite{Bernabei:2008yi}.  The
    approximate Xenon-100 exclusion region is obtained by converting
    Xenon-100's observed rate of signal candidates (3~events/100.9~days/48~kg
    between 8.4~keVnr and 44.6~keVnr~\cite{Aprile:2011hi}) into the units of
    our plots.}
    \label{fig:Zprime-nuc}
\end{figure}

As for neutrino--electron scattering, the simplest way of achieving such
enhancement is by introducing a neutrino magnetic moment.  The expected
neutrino--nucleus scattering rate for solar neutrinos with a magnetic moment at
the current upper limit is shown in figure~\ref{fig:Zprime-nuc}, curve A, for
four different target materials: germanium (used for instance in CoGeNT and
CDMS), CaWO$_4$ (used in CRESST), NaI(Tl) (used for instance in DAMA), and
xenon (used for instance in XENON-100, LUX, X-MASS, ZEPLIN, PANDA-X).  As we
can see, the effect is very small, and certainly not detectable by dark matter
experiments in the foreseeable future.

The situation is different for neutrino--nucleus scattering through the
exchange of a new light gauge boson (``dark photon'') $A'$ with mass $M_{A'}$
(see for instance the models from sections~\ref{sec:B-L}--\ref{sec:gauged-B}).
In this case, the $A'$ couplings can still be sufficiently large to allow for
substantial enhancement of the scattering rate.  Moreover, when scattering on a
heavy nucleus, a low energy neutrino cannot resolve the nuclear substructure, and
hence the scattering happens coherently on all nucleons. This leads to an
increase in the cross section proportional to the nuclear mass number $A$.

On the other hand, since nuclei are much heavier than electrons, an
$\mathcal{O}$(keV) nuclear recoil energy (above the detection threshold in a
dark matter detector) requires neutrino energies of $\mathcal{O}$(1--10~MeV),
as opposed to the $\mathcal{O}$(10~keV) required for the a detectable electron
recoil. This means that, while all solar neutrino flux components can
contribute to $\nu$--e$^-$ scattering, only the $^8$B and hep neutrinos---the
components with the highest energy cutoff and the smallest flux---can affect
the nuclear recoil signal. This leads to a reduction of 3 or 4 orders of
magnitude when compared to the $^7$Be and pp fluxes, respectively.

Moreover, the typical 4-momentum exchange $q^2 = -2 E_r m_N$ in
neutrino--nucleus scattering is much larger than in a neutrino--electron
scattering process with the same recoil energy $E_r$.  Therefore in order to
obtain substantial count rates, much higher couplings are needed in comparison
to the $\nu$--e$^-$ scenarios considered before.  Also, the larger $q^2$ means
that the transition between flat and decreasing $d\sigma/dE_r$ happens at much
larger $M_{A'}$.

In figure~\ref{fig:Zprime-nuc} (curves~B--D), we plot the $A'$-mediated
neutrino--nucleus scattering rate for three different scenarios. The first one,
curve~B, involves only the three active neutrinos, assuming that they couple
universally both to electrons and nucleons. This scenario could, for instance,
be realized in the $U(1)_{B-L}$ model from section~\ref{sec:B-L}.  To obtain a
sizeable count rate in nuclear recoils, and not violate the Borexino constraint
on neutrino--electron scattering, we need an $A'$ mass heavy enough to suppress
$\nu$--e$^-$ scattering, but not too heavy so that low energy neutrino--nucleus
scattering is still enhanced.  Curve~B in figure~\ref{fig:Zprime-nuc} is a
possible realization of this scenario which avoids all bounds to date
(see discussion in section~\ref{sec:paramspace} and figure~\ref{fig:paramspace}).
We see that, although the nuclear recoil rate in this model is higher than the
Standard Model rates, it is still at least 2--3 orders of magnitude lower than the
sensitivity of CoGeNT, CDMS, CRESST or DAMA, thus making it difficult to probe in
present and near future experiments.

Another possibility is to introduce sterile neutrinos which couple only to
quarks, but not to leptons, as is the case for instance in a model with gauged
baryon number and with sterile neutrinos charged under it (see
\cite{Pospelov:2011ha} and section~\ref{sec:gauged-B}). Since the active
neutrinos would not feel the new interaction, matter effects between the active
and sterile sectors would arise. Hence, oscillation physics should be taken
into account when computing the sterile neutrino flux at the Earth because, in
principle, it could play an important role in such a scenario.  To illustrate
the impact of oscillations, we plot curve~C in figure~\ref{fig:Zprime-nuc}, for
which we have assumed $\Delta m^2_{42} = 10^{-10}$~eV$^2$, and we have
chosen the couplings $g_\nu g_N$ such that the predicted event rates are
comparable to the ones for the $U(1)_{B-L}$ model, curve~B.\footnote{Our
choice of the mass squared difference allows to have annual modulation
compatible with DAMA or CoGeNT, see section~\ref{sec:just-so}.} Comparing
curves~B and C, we see that matter effects in this particular model change the
rates moderately, but in an energy-dependent way. We have checked that
oscillations of active neutrinos in solar and terrestrial scenarios are
unaffected for our choice of parameters.

Note that curve~C could also be realized in the model from
section~\ref{sec:kinetic-mixing}, with a gauge group $U(1)'$ under which
only the sterile neutrinos are charged, and which couples to the
Standard Model through kinetic mixing with the photon. An advantage of the
$U(1)_{B}$ scenario is that Borexino limits, as well as bounds from fixed
target experiments, atomic physics, and the anomalous magnetic moment of the
muon and the electron (see section~\ref{sec:paramspace}) can be more easily
avoided, thus opening up a generous window in the parameter space for mediator
masses from MeV to GeV and reasonably large couplings. The lack of strong
limits makes it even possible to have rates high enough to explain the current
signals in CoGeNT, CRESST and DAMA. However, since only the most energetic
$^8$B neutrinos can lead to nuclear recoils above threshold in these detectors,
and since the neutrino spectrum is steeply decreasing with energy, the event
spectra cannot be fitted very well, as can be seen in curve~D in
figure~\ref{fig:Zprime-nuc}.  For this curve, we have again assumed $\Delta
m^2_{42}\approx 10^{-10}$~eV$^2$.

%==============================================================================
\section{Annual and diurnal modulation}
\label{sec:modulation}
%==============================================================================

A generic prediction of almost all dark matter models is an annual modulation
of the dark matter interaction rate observed at the Earth. This modulation is
caused by the relative velocity of the Earth with respect to the Milky Way's
dark matter halo, which is larger in northern hemisphere summer than in winter.
In fact, the DAMA collaboration has reported a statistically significant annual
modulation in the observed event rate~\cite{Bernabei:2008yi, Bernabei:2010mq}
(see, however, \cite{Kudryavtsev:2009gd, Ralston:2010bd, Nygren:2011xu,
Blum:2011jf, Bernabei:2012wp} for a discussion of systematic effects that could
cause this modulation).  Recently, the CoGeNT collaboration has also claimed a
modulating signal~\cite{Aalseth:2011wp}, but its statistical significance is
still lower and an interpretation in terms of dark matter is
problematic~\cite{Frandsen:2011ts, Schwetz:2011xm, Farina:2011pw, Fox:2011px}
(see, however, \cite{Hooper:2011hd, Belli:2011kw} which come to different
conclusions). Both DAMA and CoGeNT observe the maximum count rate during the
northern hemisphere summer or spring months (early June for DAMA and mid-April
for CoGeNT).  Other experiments have not yet published searches for annual
modulation, but once an experiment has collected a sufficient number of
candidate events, such a search would be the logical next step. Another
potential ``smoking gun'' signature of dark matter is diurnal modulation,
caused by the changes in relative velocity between the detector and the dark
matter halo during the day. This modulation is predicted to be much smaller
than the annual one, but it is nevertheless being searched
for~\cite{Bernabei:1999pt, Fox:2011px}, though with negative results so far. A
significant improvement in the sensitivity to daily modulation is expected in
dark matter detectors with directional sensitivity~\cite{Ahlen:2009ev}.

Because of the DAMA and CoGeNT signals, but also to scrutinize the robustness
of the dark matter interpretation of \emph{any} annual or diurnal modulation
signal in the future, it is important to investigate alternative sources of
temporal modulation in a dark matter detector. In this section we investigate
modulation signals that could arise from neutrino physics beyond the Standard
Model, and we discuss if and how these signals can be distinguished from a dark
matter signal.

%------------------------------------------------------------------------------
\subsection{The Earth--Sun distance}
%------------------------------------------------------------------------------

Any signal whose source is an isotropic particle flux from the Sun is expected
to show annual modulation due to the fact that the flux decreases with the
square of the Earth--Sun distance, and that the Earth's orbit is slightly
elliptical. At its closest point, the perihelion, around January 3rd, the
Earth--Sun distance is 0.983~AU, versus 1.017~AU at the aphelion around July
4th. Hence, a solar neutrino signal will be modulated with an amplitude of
about 3\%, with the maximum count rate expected in early January.  The expected
phase would thus be \emph{opposite} to the one observed in DAMA, but it is
important to note that, depending on the dark matter mass and velocity
distribution, dark matter can also lead to modulating signals peaking in
winter~\cite{Freese:1987wuFig8}. Thus, a neutrino signal modulating
predominantly because of the ellipticity of the Earth's orbit cannot explain
the DAMA signal, but could be confused with other dark matter signals.

In the following, we will see that there may be additional sources of
modulation that can potentially be large enough to overcompensate the
Earth--Sun distance effect and thus reverse the sign of the modulation, bringing
it in line with the DAMA observation.

%------------------------------------------------------------------------------
\subsection{Annual modulation from neutrino oscillations in vacuum}
\label{sec:just-so}
%------------------------------------------------------------------------------

\begin{figure}
  \begin{center}
    \includegraphics[width=12cm]{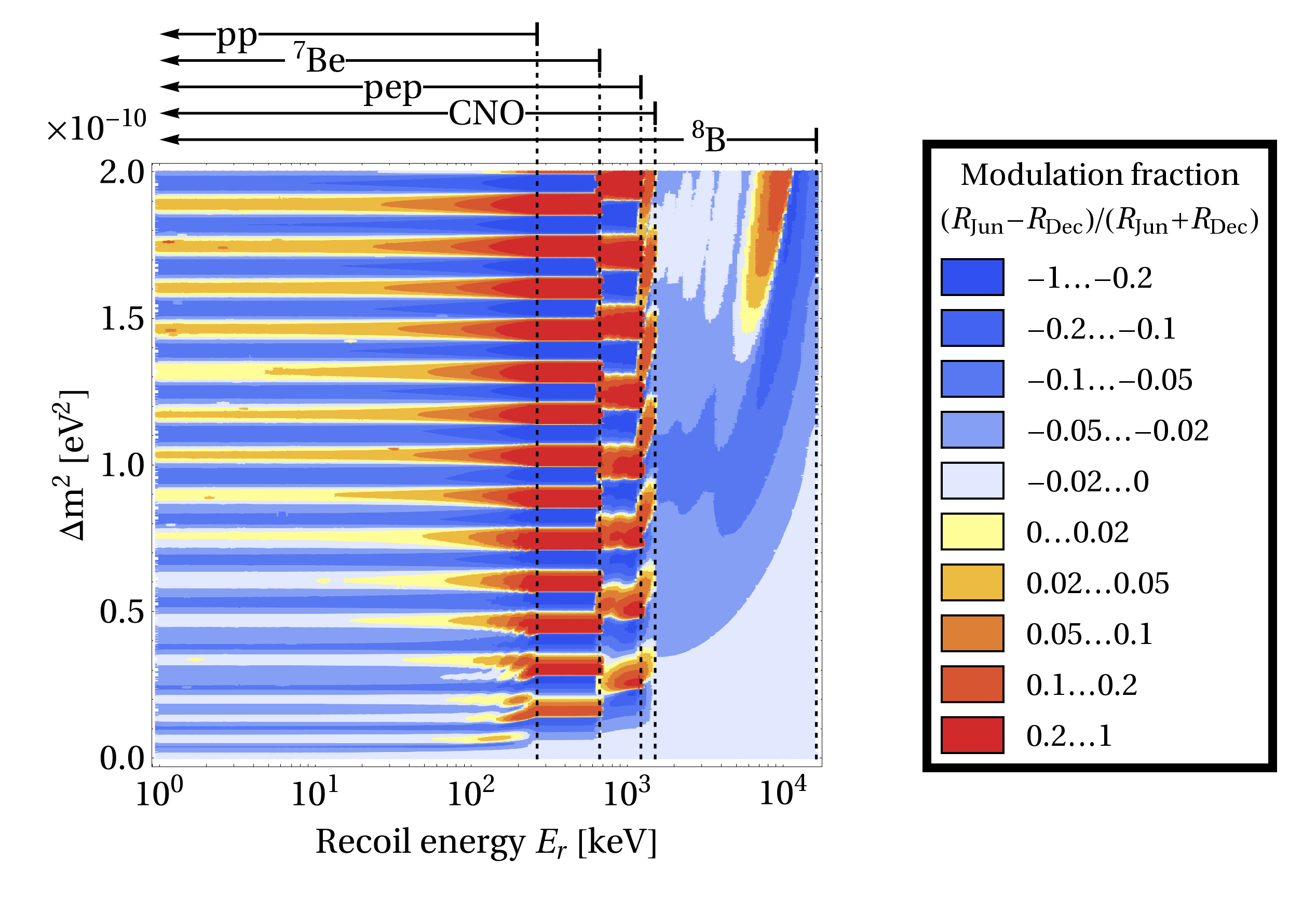}
  \end{center}
  \caption{Relative annual modulation fraction $(R_{\rm Jun} - R_{\rm Dec})
    / (R_{\rm Jun} + R_{\rm Dec})$ of the rate of $A'$-mediated sterile
    neutrino--electron scattering as a function of the recoil energy $E_r$ and
    the mass squared difference between active and sterile neutrinos.  For
    simplicity, we use a two-flavor vacuum oscillation framework here. One can
    clearly distinguish five different regimes, in which the rate is dominated
    by pp neutrinos, \iso{Be}{7} neutrinos, pep neutrinos, CNO neutrinos, and
    \iso{B}{8} neutrinos, respectively.}
  \label{fig:just-so}
\end{figure}

A simple way of overcompensating for the Earth--Sun distance effect is to
assume that the anomalous signal we are interested in is exclusively due to new
light sterile neutrinos (see e.g.\ the $U(1)'$ model from
section~\ref{sec:kinetic-mixing} or the $U(1)_B$ model from
section~\ref{sec:gauged-B}), and that these sterile neutrinos mix with the
active neutrinos in such a way that the oscillation probability is increasing
with increasing Earth--Sun distance.  Then, oscillations would modify the
simple $L^{-2}$ scaling of the sterile neutrino flux with distance $L$, and in
particular the sign of the modulation can be reversed~\cite{Pospelov:2011ha}.
This is illustrated in figure~\ref{fig:just-so}, where we plot the expected
relative annual modulation amplitude, $(R_{\rm Jun} - R_{\rm Dec}) / (R_{\rm
Jun} + R_{\rm Dec})$ for sterile neutrino--electron scattering as a function of
the electron recoil energy $E_r$ and the mass squared difference $\Delta m^2$
between the mostly active and mostly sterile mass eigenstates. (Here, $R$
stands for the count rate differential in $E_r$, and the subscripts ``Jun'' and
``Dec'' indicate the count rate at the northern hemisphere summer and winter
solstice, respectively.) Note that, in producing figure~\ref{fig:just-so}, we
have worked in a simple two-flavor oscillation framework, a situation which
could be realized in a 4-neutrino model, where the mostly sterile mass
eigenstate $\nu_4$ mixes predominantly with only one of the active mass
eigenstates, say $\nu_2$. We have moreover assumed that the conversion of
active neutrinos into sterile neutrinos proceeds through vacuum oscillations.
This is realized for instance in the $U(1)'$ model with kinetic mixing from
section~\ref{sec:kinetic-mixing}, in which no non-standard matter effects arise
due to the zero net electric charge of the Sun.  In models where the
two-flavor/vacuum approximation is not justified, the phenomenology could be
much richer and much more complicated.

We see from figure~\ref{fig:just-so} that a sizeable annual modulation fraction
can be achieved without undue fine-tuning of $\Delta m^2$ \emph{if} the
oscillation length $L_{\rm osc} = 4\pi E_\nu / \Delta m^2$ at the relevant
neutrino energies between $\text{few} \times 10$~keV and $\sim 15$~MeV is
smaller but still comparable to the Earth--Sun distance.  Such a situation
would be reminiscent of the now excluded ``just-so'' solution to the solar
neutrino problem, where it was also hypothesized that the oscillation length of
solar neutrinos could be comparable to 1~AU. It is easy to distinguish four
different energy regimes in figure~\ref{fig:just-so}, in which the scattering
rate is dominated by pp neutrinos, \iso{Be}{7} neutrinos, pep neutrinos, CNO
neutrinos, and \iso{B}{8} neutrinos, respectively. In each of these regimes,
the modulation fraction is determined mostly by the oscillation length at the
peak energy of the corresponding neutrino flux, as long as the recoil energy is
large enough for this peak energy to be kinematically accessible.

Note that for $\Delta m^2$ values larger than the ones shown in
figure~\ref{fig:just-so}, the impact of oscillations will fade away once the
oscillation length at the relevant neutrino energies becomes smaller than the
diameter of the neutrino production region in the Sun.  For pp neutrinos, which
are produced at radii $r < 0.2 R_\odot$ in the Sun and whose flux peaks at
around 300~keV, this happens for $\Delta m^2 \gtrsim 3 \times 10^{-9}$~eV$^2$,
whereas for \iso{B}{8} neutrinos, which are produced at $r < 0.1 R_\odot$ and
whose flux peaks at around 6~MeV, wash-out effects become relevant only for
$\Delta m^2 \gtrsim 10^{-7}$~eV$^2$~\cite{Bahcall:1989}.

Note also that for $\Delta m^2 \sim 10^{-10}$~eV$^2$, the Standard Model MSW
potential (or the $A'$-induced matter potential in the case of the $U(1)_B$
models from section~\ref{sec:gauged-B}) will suppress the mixing of electron
neutrinos with sterile neutrinos at the center of the Sun. As the neutrinos
propagate out, they can pass through an MSW resonance (if $\Delta m^2$ has the
appropriate sign) and acquire an admixture of the mostly sterile mass
eigenstate $\nu_4$ because the resonance transition is non-adibatic, for such
small $\Delta m^2$. (This can be seen from equation~\eqref{eq:adiabaticity}
below.)

%------------------------------------------------------------------------------
\subsection{Diurnal and annual modulation from Earth matter effects}
\label{sec:earth-matter-modulation}
%------------------------------------------------------------------------------

Another mechanism by which solar neutrino signals can modulate with time is
Mikheyev-Smirnov-Wolfenstein (MSW) type matter effects in the Earth. It is well
known (see for instance reference~\cite{Akhmedov:2004rq} and references
therein) that even in the standard three-flavor oscillation framework
matter-enhanced $\nu_\mu, \nu_\tau \to \nu_e$ oscillations of solar neutrinos
inside the Earth can lead to a slightly enhanced $\nu_e$ flux during the night,
when solar neutrinos have to traverse the Earth before reaching a detector.
This leads to diurnal modulation of the $\nu_e$ detection rate, and, since
nights are longer in winter than in summer, it  also leads to annual
modulation.  In the standard framework, the day--night asymmetry is predicted
to be very small, on the few per cent level, but we will argue here that it can
be sizeable in the new physics sector.

Consider, for instance, a scenario based on the model from
section~\ref{sec:gauged-B} (a $U(1)_B$ gauge boson), but with \emph{two}
sterile neutrinos, weakly mixed with the active ones. We assume that one of the
sterile neutrino flavors, say $\nu_{s1}$, is charged under the new gauge group,
whereas the other, $\nu_{s2}$, is not.  Thus, only $\nu_{s1}$ can be observed
in a detector.  To simplify the discussion, we also assume both of the mostly
sterile mass eigenstates to be heavy enough for them to be never produced in
coherent superposition with the mostly active mass eigenstates, so that
oscillations among the active flavors are fully decoupled from oscillations
among the sterile flavors. We do, however, assume the mass splitting $\Delta
m^2$ between the sterile neutrinos to be sufficiently small for oscillations
among them to occur. We can then discuss these oscillations in a simple
two-flavor framework.

Using the well-known formalism of neutrino oscillations in matter, it is
straightforward to show that the probability for a solar neutrino to arrive
in a terrestrial detector in the $\nu_{s1}$ flavor eigenstate is given by
(see appendix)~\cite{Akhmedov:2004rq}
\begin{align}
  P(\nu_e \to \nu_{s1}) = |U_{e4}^\odot|^2 \cos^2\theta
    + |U_{e5}^\odot|^2 \sin^2\theta
    + (|U_{e5}^\odot|^2 - |U_{e4}^\odot|^2)
        \sin^2 2\theta \frac{2 E V_{A'}^\oplus}{\omega^2 \Delta m^2}
        \sin^2 \frac{\omega \Delta m^2 L^\oplus}{4 E} \,.
  \label{eq:solar-prob}
\end{align}
Here, $U_{e4}^\odot$ and $U_{e5}^\odot$ are elements of the effective leptonic
mixing matrix in matter at the core of the Sun, $\theta$ is the vacuum mixing
angle between $\nu_{s1}$ and $\nu_{s2}$, $E$ is the neutrino energy, $L^\oplus$
is the distance the neutrinos travel inside the Earth, and $V_{A'}^\oplus$ is
the $A'$-mediated MSW matter potential, equation~\eqref{eq:V-condition}, in the
Earth, which by assumption affects only the flavor eigenstate $\nu_{s1}$, but
not $\nu_{s2}$.  For simplicity, we take $V_{A'}^\oplus$ to be constant
throughout the Earth.  We have also introduced the abbreviation
\begin{align}
  \omega \equiv \sqrt{\sin^2 2\theta + (\cos 2\theta - 2 E V_{A'}^\oplus / \Delta m^2 )^2} \,.
  \label{eq:omega}
\end{align}
Note that equation~\eqref{eq:solar-prob} is valid only in the case where flavor
transitions in the Sun are fully adiabatic, which requires~\cite{Akhmedov:1999uz}
\begin{align}
  \frac{\Delta m^2 \sin^2 2\theta}{2E\cos 2\theta} \gg
    \bigg|\frac{\dot{V}_{A'}^\odot}{V_{A'}^\odot}\bigg| \,,
  \label{eq:adiabaticity}
\end{align}
where $\dot{V}_{A'}^\odot$ denotes the derivative of the non-constant matter
potential inside the Sun with respect to distance from the center.  In the
following, we assume this adiabaticity condition to be fulfilled.
Note also that, for sufficiently heavy sterile neutrinos, $V_{A'}^\odot \ll
|\Delta m^2_{4i}|/2E$, hence the active--sterile mixing angles in matter and vacuum
are almost identical: $U_{e4}^\odot \sim U_{e4}$, $U_{e5}^\odot \sim U_{e5}$.

For oscillation lengths much shorter than the diameter of the Earth, a detector
with limited statistics and limited energy resolution will not be able to
resolve the individual oscillation peaks, but will only see their average
effect.  In this case we can make the replacement $\sin^2 (\omega \Delta m^2
L^\oplus / 4E) \to 1/2$ in equation~\eqref{eq:solar-prob}. We see that, if the
resonance condition
\begin{align}
  2 E V_{A'}^\oplus \simeq \Delta m^2 \cos 2\theta
  \label{eq:A'-resonance}
\end{align}
is fulfilled, the oscillation amplitude inside the Earth can be very large even
if the vacuum mixing angle $\theta$ is small.  If $|U_{e4}^\odot|^2 >
|U_{e5}^\odot|^2$, the probability for detecting a $\nu_{s1}$ is larger during
the day than it is at night.  As discussed above, this diurnal modulation also
leads to an annual modulation of the daily average count rate. The length of
the day is larger in summer than in winter, therefore, for $|U_{e4}^\odot|^2 >
|U_{e5}^\odot|^2$ ($|U_{e4}^\odot|^2 < |U_{e5}^\odot|^2$), the average daily
count rate is also larger (smaller) in summer than in winter.  Note that this
type of annual modulation cannot be invoked to explain the modulation signals
observed in DAMA~\cite{Bernabei:2010mq} and CoGeNT~\cite{Aalseth:2011wp} since
neither of these experiments has observed the accompanying (and stronger)
diurnal modulation~\cite{Bernabei:1999pt, Fox:2011qd}.

It is also very important to recall that a non-standard matter potential in the
Sun ($V_{A'}^\odot$) and in the Earth ($V_{A'}^\oplus$) for sterile neutrinos is
only generated in models in which the Sun and the Earth are not neutral under
the gauge group that couples Standard Model particles to the sterile neutrinos.
Therefore, the modulation mechanism discussed here will not be effective in
models in which the new gauge boson couples to the Standard Model only through
kinetic mixing.

It is useful to rewrite the resonance condition~\eqref{eq:A'-resonance} as
\begin{align}
  L_{\rm osc} V_{A'}^\oplus = 2\pi \cos 2\theta \,,
  \label{eq:Losc-condition}
\end{align}
with the oscillation length $L_{\rm osc} = 4\pi E / \Delta m^2$. From this
expression we see that for large $V_{A'}^\oplus$, i.e.\ relatively strong
coupling between $\nu_{s1}$ and ordinary matter the resonance condition can
only be fulfilled if the oscillation length is very small, i.e.\ for very low
$E$ or relatively large $\Delta m^2$. If $L_{\rm osc} \lesssim 1$~km, matter
effects are important even during daytime because detectors are typically
located $\gtrsim 1$~km underground.  Setting $L_{\rm osc} = 1$~km in
equation~\eqref{eq:Losc-condition}, we find that this happens for
\begin{align}
  \frac{g_e g_\nu}{M_{A'}^2} \gtrsim 0.18\ \text{GeV}^{-2} \,.
  \label{eq:gegnu-condition}
\end{align}
This should be compared to the corresponding Standard Model quantity $g^2 / 4
M_W^2 = 8.2 \times 10^{-6}$~GeV$^{-2}$.

It is amusing to note that a strong day--night asymmetry combined with the
effects of the varying Earth--Sun distance could conspire to give a modulation
phase that is very dark matter-like. The amplitude of annual modulation from a
day--night asymmetry is at most on the order of 20\% for a detector at
mid-latitudes, and the peak of such a modulation coincides with the summer
solstice around June 21st.  The Earth--Sun distance effect is weaker and has a
nearly opposite phase, with a minimum on July 4th. The combined effect is a
modulation with a maximum total flux occurring on a date earlier than June 21st
and closer to early June, which is the canonical peak of the dark matter
signal. This scenario may be easily distinguished from a dark matter signal
thanks to the strong daily modulation which it predicts (and which has not been
seen in DAMA and CoGeNT).

%------------------------------------------------------------------------------
\subsection{Zenith angle dependence of Earth matter effects}
%------------------------------------------------------------------------------

\begin{figure}
  \begin{center}
    \includegraphics[width=8cm]{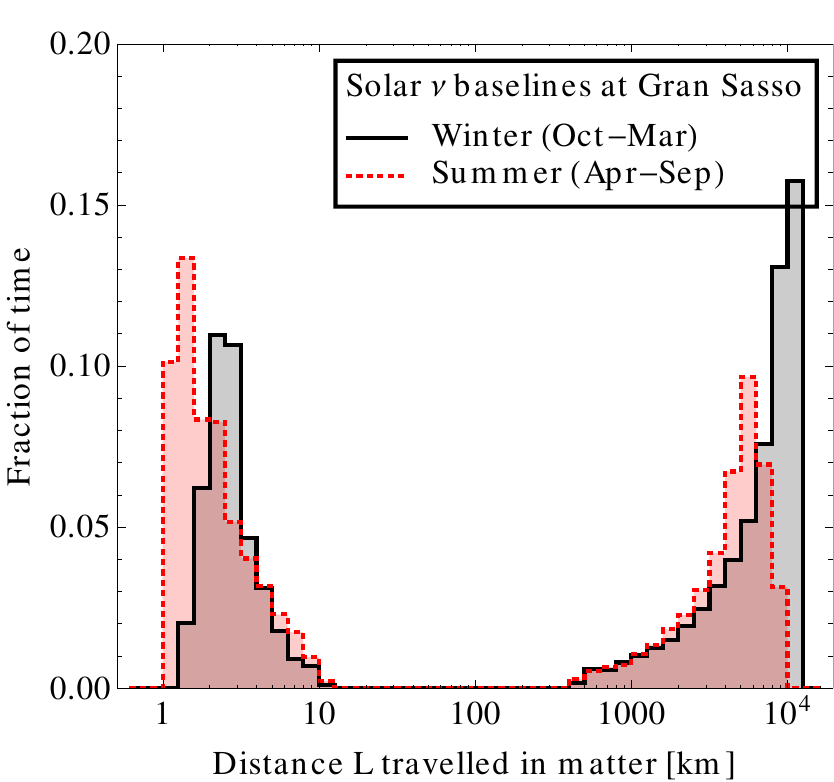}
  \end{center}
  \caption{The distribution of the amount of Earth matter a
      solar neutrino has to travel through in winter (October--March,
      black histogram) compared to summer (April--September, red
      histogram) in order to reach a detector in the Gran Sasso
      Laboratory.  To obtain this plot, we have computed the thickness
      of the rock around the laboratory as a function of
      the zenith and azimuthal angles (including effects of local
      topography), and have weighted the result by the fraction of
      time the detector sees the Sun in any given direction.  A very
      large daily and annual modulation of a sterile neutrino signal
      may be achieved in models with oscillation lengths around a few
      kilometers or around several thousand kilometers.}
  \label{fig:overburden}
\end{figure}

In the previous section, we have seen that a day--night asymmetry in the sterile
neutrino interaction rate in a detector can be generated in models where the
oscillation length $L_{\rm osc}$ for oscillations among different sterile
neutrino flavors is much smaller than the diameter of the Earth ($\sim
12\,700$~km), but much bigger than the detector's rock overburden (typically of
order one kilometer).  Phenomenologically very interesting (though extremely
fine-tuned) models can be constructed by choosing the mass splitting between
the new neutrino mass eigenstates such that $L_{\rm osc}$ is close to either
end of this range.  This is because the daily averaged distance a solar
neutrino must travel inside rock before reaching a detector modulates annually
due to the different range of zenith angles under which the detector sees the
Sun at different times of the year. Away from the equator, the Sun is higher in
the sky at noon in summer as compared to winter. Similarly, in winter the Sun
drops lower below the horizon during the night.  In figure~\ref{fig:overburden}
we show the distribution of neutrino path lengths in matter for solar neutrinos
on their way to the Gran Sasso laboratory (where, for instance, the DAMA,
CRESST and XENON-100 experiments are located). In the plot we distinguish
between the summer months (red histogram) and the winter months (black
histogram).  We see that both the average path length in matter during daytime
and during nighttime varies by an $\mathcal{O}(1)$ factor between summer and
winter.  Thus, if $L_{\rm osc}$ is around one kilometer or around several
thousand kilometers, the oscillation probability can modulate significantly
during the year (and also during a day).  It is amusing to note that in the
case $L_{\rm osc} \sim 1$~km, the precise modulation pattern depends on the
latitude of the laboratory, on the detailed topography of the landscape above
the detector, and possibly even on the location of the detector within the
laboratory.  To compute figure~\ref{fig:overburden}, we have used topographical
data of the Gran Sasso region~\cite{ASTER-GDEM} to determine the rock thickness
$d(\Theta, \Phi)$ above the Gran Sasso Laboratory
(42$^\circ$~27'~North, 13$^\circ$~34'~East, 963~m above sea
level~\cite{Bellotti:1989pa}) as a function of the zenith angle $\Theta$ and
the azimuth angle $\Phi$.  For zenith angles below the horizon, we have
approximated the Earth as a simple sphere. We have then weighted $d(\Theta,
\Phi)$ by the fraction of time the Sun spends at any given point $(\Theta,
\Phi)$ in the sky (or below the horizon) during the time periods indicated in
the legend.  All computations were done in Mathematica~8.

%------------------------------------------------------------------------------
\subsection{Diurnal and annual modulation from neutrino absorption in the Earth}
%------------------------------------------------------------------------------

In models that feature sterile neutrinos with sufficiently large
couplings to ordinary matter, the sterile neutrinos' scattering cross
section can be so large that their mean free path becomes less than
the diameter of the Earth.  (For constraints on such models and a
discussion of the allowed parameter space see
section~\ref{sec:paramspace} below.) At night, when they have to
travel through a substantial amount of matter before reaching a
detector, the sterile neutrinos would thus loose all their kinetic
energy and become undetectable, whereas during daytime, they could
reach the detector unimpeded. This can lead to a very strong daily
modulation of the experimental event rate, and due to the different
length of day in summer compared to winter, also to annual modulation
peaking in summer.

%------------------------------------------------------------------------------
\subsection{Modulation from direction-dependent quenching factors}
%------------------------------------------------------------------------------

Finally, there is the possibility that temporal modulation of a neutrino
scattering signal is induced by direction-dependent solid state effects in a
target crystal. This source of modulation is especially interesting for signals
which originate from a particular direction such as the Sun, as opposed to
signals originating from dark matter which are roughly isotropic, with only a
small direction dependence in the velocity spectrum due to the Earth's motion
with respect to the dark matter halo of the Milky Way. (This directionality in
the dark matter velocity distribution as seen from the Earth is usually
referred to as the ``WIMP wind'').

It is well known that the response of a solid state detector to nuclear recoils
can be very sensitive to the direction in which the recoil nucleus is traveling
with respect to the crystal axes (see for instance~\cite{Bozorgnia:2010xy,
Bozorgnia:2010ax, Bozorgnia:2010er, Bozorgnia:2010zc, Bozorgnia:2011tk,
Bozorgnia:2011ve}).  In particular, if the initial momentum of the recoiling
nucleus is aligned with one of the crystal planes, it is likely to bump into
its nearest neighbors, and most of its energy will be converted into phonons.
On the other hand, if the recoiling nucleus enters the space between crystal
planes and travels along this ``channel'', it will mostly scatter on electrons,
so that a larger fraction of the recoil energy is converted into electronic
excitations. In many detectors, only electronic excitations can be detected, so
that for these ``channeled'' events, the ratio between the visible energy and
the actually deposited energy (the quenching factor) is larger.  The magnitude
of blocking and channeling effects is strongly dependent on the target
material, the nuclear recoil energy and the temperature at which the detector
is operated. In typical dark matter detectors, at most a few per cent of
nuclear recoils with a given energy and direction can be channeled, and
\emph{typical} channeling fractions are 1--3 order of magnitude
smaller~\cite{Bozorgnia:2010xy, Bozorgnia:2010ax, Bozorgnia:2010er,
Bozorgnia:2010zc, Bozorgnia:2011tk}.

Detectors like CDMS or CRESST, which use superconducting phase transition
thermometers to measure directly the total deposited energy, are not strongly
affected by channeling, and only their background rejection efficiency might
change if the fraction of energy going into electronic excitations
(scintillation and ionization) varies.  On the other hand, experiments like
DAMA or CoGeNT, which rely exclusively on electronic signals for their energy
measurement, could be strongly affected by direction-dependent quenching
factors.

\begin{figure}[t]
  \begin{center}
    \includegraphics[width=\textwidth]{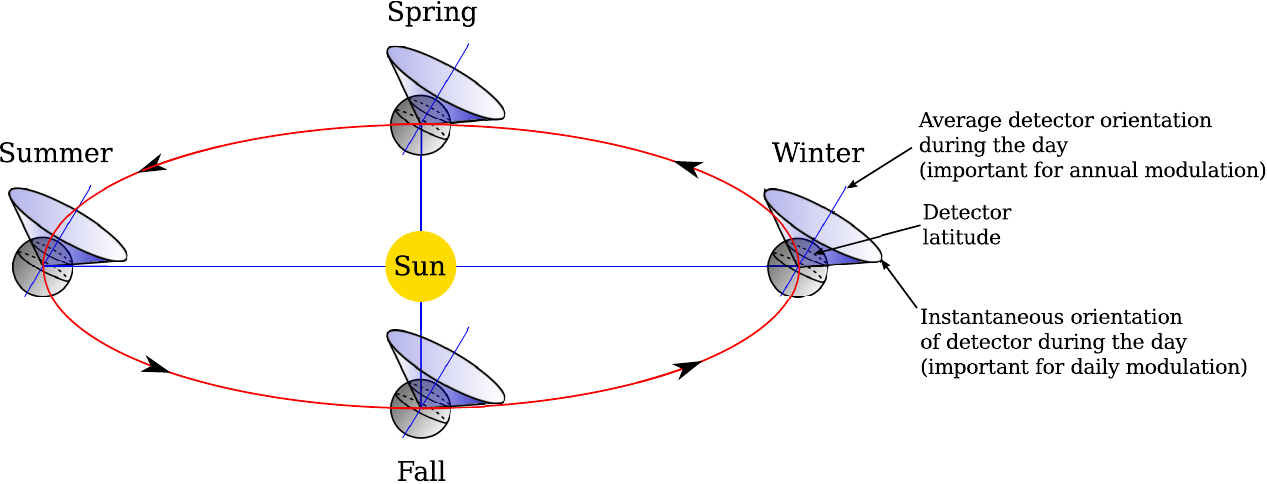}
  \end{center}
  \caption{Illustration of the relative orientation of a detector with respect
    to the solar neutrino flux throughout the year. In blue we show the cones
    which the detector's normal axis traces out during a day, and in red we
    show the Earth's trajectory around the Sun.  As explained in the text, a
    detector whose detection efficiency depends on the direction of the
    incoming particles can observe both diurnal and annual modulation in the
    solar neutrino signal, and possibly also several higher harmonics of these
    fundamental frequencies.}
  \label{fig:directional-sensitivity}
\end{figure}

In these experiments, the reconstructed recoil energy spectrum, and thus also
the total number of events above threshold, would change as a function of the
recoil direction. For conventional signals from the scattering of heavy dark
matter particles on nuclei, the rotation of the Earth implies that at different
times during the day, the detector sees the ``WIMP wind'' under different
angles, so that one would expect a small modulation in the detection rate
during the sidereal day (not the solar day).  Direction dependent detection
efficiencies would, however, not contribute to annual modulation because the
orientation of the Earth's axis relative to the WIMP wind does not change
during the year. This can be understood from
figure~\ref{fig:directional-sensitivity}, in which the blue cones illustrate
the trajectory which the detector's normal axis traces during a day, and the
red ellipse depicts the Earth's orbit around the Sun. 
%We see that 

The situation is different for signals induced by particles coming from the
Sun, such as neutrinos. In this case, direction-dependent detection
efficiencies will lead to modulation synchronized with the \emph{solar} day
(not the sidereal day) and with the solar year. The exact spectrum of
modulation frequencies depends on the specific target material, in particular
on the orientation of the preferred directions for channeled nuclear recoils in
this material. The important time scales for modulation in this scenario will
almost certainly be daily and annually, but semi-annual or other periods are
also possible, depending on the symmetries and orientation of the crystal in
question.

Modulation signals from direction-dependent quenching factors may be easily
identified by rotating the detector and collecting more data.

%==============================================================================
\section{The parameter space for light gauge bosons and new constraints}
\label{sec:paramspace}
%==============================================================================

One of the central ingredients of the models discussed in the previous sections
is a new, light and feebly coupled $U(1)'$ gauge boson $A'$, the dark photon.
The possible existence of such a particle has been considered in the literature
in many contexts (unrelated to neutrinos), and strong bounds on the $A'$ mass
and couplings have been derived. Therefore, in order to fully assess the
viability of our models, we need to make contact with these experimental
constraints.  In order to keep the discussion brief we will in most cases
simply present the existing bounds and refer the reader to the literature for
further details on the physics behind them.  We will consider the models
discussed in sections~\ref{sec:B-L}--\ref{sec:gauged-B}:
\begin{enumerate}[\hspace{1.5em}\bf A]
  \item A $U(1)_{B-L}$ gauge boson with vector couplings to fermions
    \label{model:B-L}

  \item A $U(1)'$ gauge boson kinetically mixed with the photon, and
    with no other couplings to Standard Model particles
    \label{model:km}

  \item A $U(1)_{B}$ gauge boson (gauged baryon number)
    \label{model:B}
\end{enumerate}
As we have seen in sections~\ref{sec:electron-recoil} and
\ref{sec:nuclear-recoil}, model~\ref{model:B-L} can substantially enhance even
the scattering rates of active neutrinos. In model~\ref{model:km}, on the other
hand, the dark photon does not couple to electrically neutral Standard Model
particles such as neutrinos, so this model can lead to large neutrino signals
in dark matter detectors only if, in addition to the dark photon, there are
sterile neutrinos that carry a $U(1)'$ charge. In this case, the dark photon
can induce sterile neutrino--electron and sterile neutrino--nucleus scattering,
with the latter effect being subdominant.  See
sections~\ref{sec:electron-recoil} and \ref{sec:nuclear-recoil} for details on
the phenomenology. Finally, model~\ref{model:B} can lead to an enhancement of
only the neutrino--nucleus scattering rate, while leaving neutrino--electron
scattering largely unchanged (except for loop-induced effects), provided that
there are ``baryonic'' sterile neutrinos charged under $U(1)_B$.

In the literature, constraints on light gauge bosons are usually presented in
the context of $U(1)'$ bosons coupled only through kinetic mixing
(model~\ref{model:km}), but for many of the relevant processes (see below), a
scenario with a $U(1)_{B-L}$ gauge boson is equivalent to such a model. In
fact, for any process involving only $A'$ couplings to electrons and protons
(but not neutrons or neutrinos), whose electric charges happen to be identical
to their $B-L$ charges, a $U(1)_{B-L}$ gauge boson with coupling $g_{B-L}$ and
mass $M_{A'}$ is equivalent to a kinetically mixed dark photon,
equation~\eqref{eq:L-light-steriles}, with kinetic mixing parameter $\epsilon =
g_{B-L} / e \times [1 + g_{B-L}^2 / e^2]^{-1/2}$ and mass $M_{A'}^2 \times (1 +
g_{B-L}^2 / e^2)^{-1}$. At the formal level, this equivalence can be
demonstrated by transforming the $B-L$ model according to $A \to A - g_{B-L} /
e \times A'$, and then rescaling the $A'$ field. (Here, $A$ is the Standard
Model photon.)  For $U(1)_B$ gauge bosons, model~\ref{model:B}, many of the
existing constraints do not apply at all (see below).

Where applicable, we will also comment on variations of
models~\ref{model:B-L}--\ref{model:B}, in particular scenarios with sterile
neutrinos $\nu_s$ charged under $U(1)'$, where $A' \to \bar\nu_s \nu_s$ decays
can modify the phenomenology. We will also mention ``chameleon'' models, i.e.\
models in which the $A'$ mass changes as a function of the background matter
density~\cite{Nelson:2007yq, Nelson:2008tn, Feldman:2006wg}, which can also
change or eliminate bounds.

\begin{figure}
  \begin{center}
    \includegraphics[width=\textwidth]{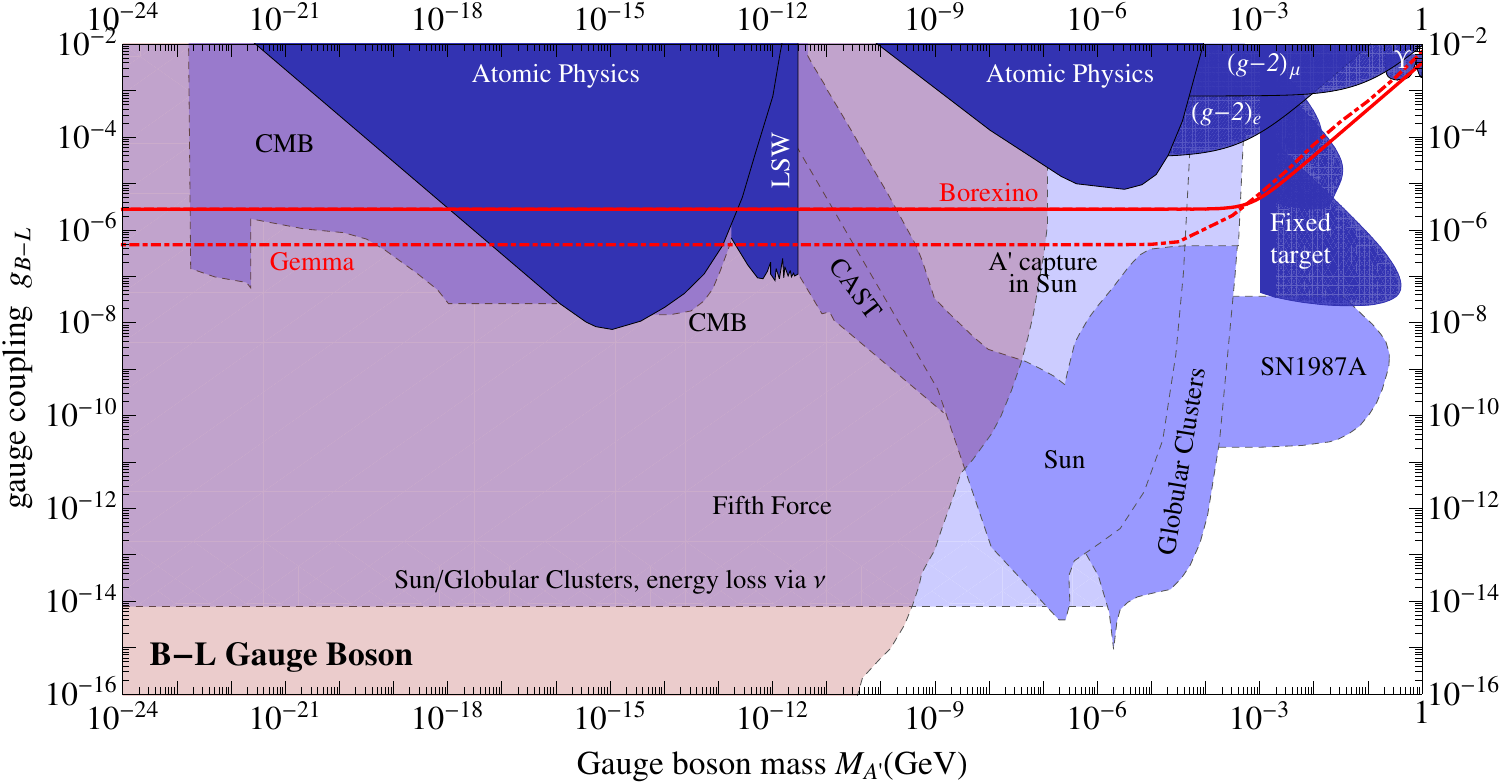} \\[0.7cm]
    \includegraphics[width=\textwidth]{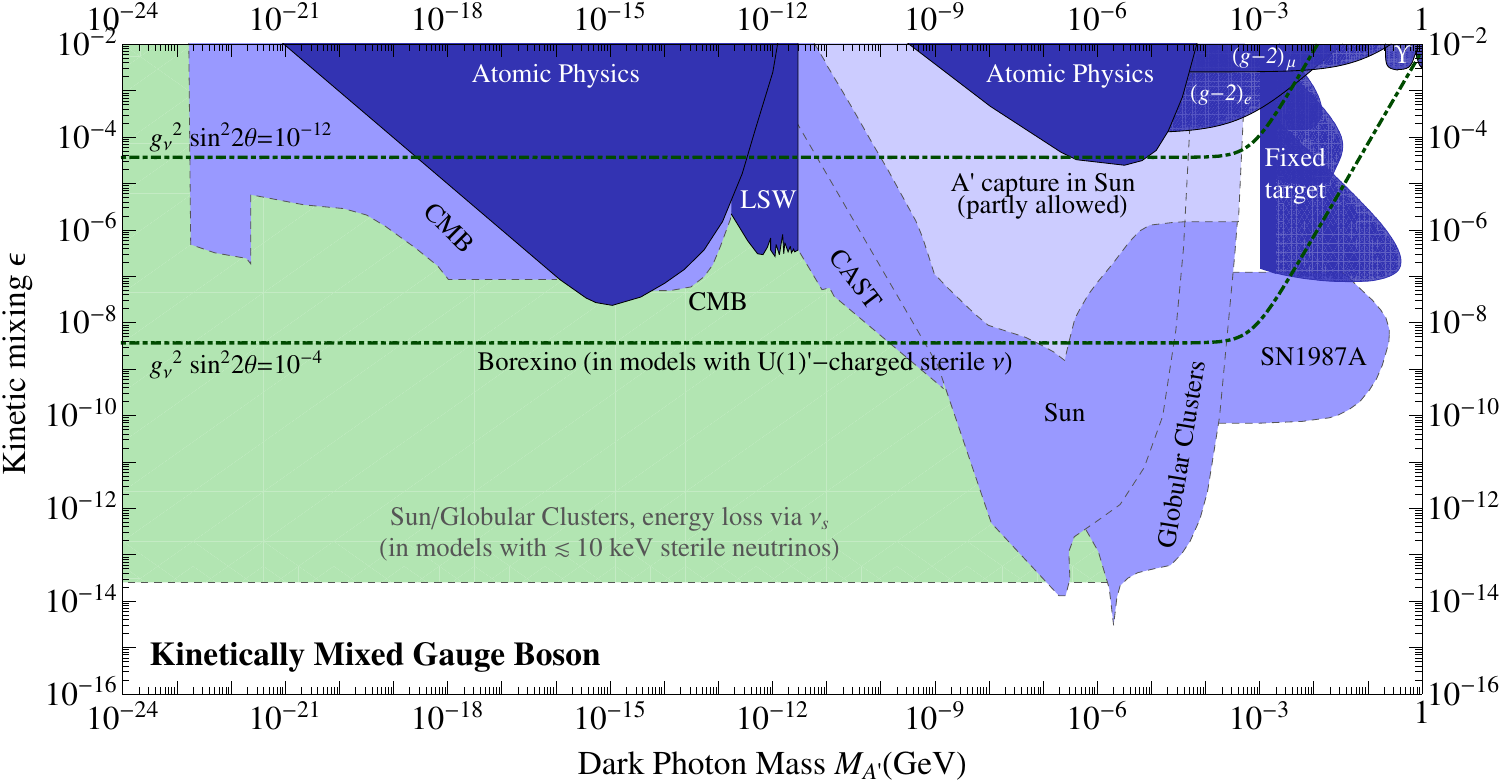}
  \end{center}
  \caption{\emph{Top:} Constraints on a $U(1)_{B-L}$ gauge boson
    (model~\ref{model:B-L}, section~\ref{sec:B-L}) with coupling $g_{B-L}$ and
    mass $M_{A'}$.  \emph{Bottom:} Constraints on light $A'$ gauge bosons
    kinetically mixed with the photon (model~\ref{model:km},
    section~\ref{sec:kinetic-mixing}) as a function of the $A'$ mass and the
    kinetic mixing parameter~$\epsilon$.  The various bounds are briefly
    explained in the text.  Limits shown in dark blue are the ones that
    \emph{cannot} be evaded even if the mass of the $A'$ depends strongly on
    the local matter density (so-called chameleon effects~\cite{Nelson:2007yq,
    Nelson:2008tn, Feldman:2006wg}). The red exclusion regions apply only to
    the $U(1)_{B-L}$ models (\ref{model:B-L}), but not models with only kinetic
    mixing (model~\ref{model:km}). The green exclusion regions apply only if
    the model contains sterile neutrinos directly charged under $U(1)'$.  Most
    limits are taken from the compilation in~\cite{Jaeckel:2010ni}, see text
    for further references. To our knowledge, the Borexino and GEMMA limits
    shown here have not been discussed before.}
  \label{fig:paramspace}
\end{figure}

The various constraints on $U(1)_{B-L}$ gauge bosons are shown in the top panel
of figure~\ref{fig:paramspace}, and the constraints on light gauge bosons
kinetically mixed with the photon are shown in the bottom panel.  Constraints
on all three models considered here are also summarized in
table~\ref{tab:constraints}.  To obtain the top panel of
figure~\ref{fig:paramspace}, we have translated the limits on kinetic mixing
from~\cite{Jaeckel:2010ni,Dent:2012mx} into limits on a $B-L$ gauge coupling
$g_{B-L}$ by replacing $\epsilon$ with $g_{B-L} / e$, where $e$ is the unit
charge. (In the case of the constraint from $\Upsilon$ decays, the
transformation is $\epsilon \to g_{B-L} / 2 e$ because the electric charge of
the bottom quark is twice its $B-L$ charge.) It is important to note that there
are additional constraints in the $U(1)_{B-L}$ case (model~\ref{model:B-L})
compared to model~\ref{model:km} (kinetic mixing), namely from Borexino and
from long-range force searches~\cite{Adelberger:2006dh, Adelberger:2009zz,
Bordag:2001qi, Bordag:2009} (see below for details).  The latter apply also to
the $U(1)_B$ model (model~\ref{model:B}).

The bounds in the plot are color coded: Regions colored in dark blue represent
the most robust limits from laboratory experiments. In particular, these limits
are still valid if the new light gauge boson is subject to strong chameleon
effects.  Limits in lighter shades of blue are sensitive and may be evaded with
chameleon effects (the lightest shades of blue are reserved for limits that
have not been worked out in detail and are less robust. Limits shown in red
apply to the $U(1)_{B-L}$ model (\ref{model:B-L}), but not models in which the
$A'$ couples to Standard Model particles only via kinetic mixing
(model~\ref{model:km}).  Green limits apply only if there are, in addition to
the $A'$ gauge boson, sterile neutrinos charged under $U(1)'$ with the coupling
constant and mixing angle indicated in the plot.

The individual constraints, which are also summarized in
table~\ref{tab:constraints}, are:
\begin{enumerate}
  \item {\bf Anomalous magnetic moment of the electron and the muon
    (``$(g-2)_e$'', ``$(g-2)_\mu$'').} A new gauge boson coupling to leptons
    (Models \ref{model:B-L} and \ref{model:km}, but not \ref{model:B}) will
    contribute to the magnetic moment of the electron and the muon at the one
    loop level, see for example~\cite{Pospelov:2008zw}. Note that a model with
    parameters just below the $(g-2)_\mu$ exclusion line in
    figure~\ref{fig:paramspace} could explain the currently observed deviation
    of $(g-2)_\mu$ from its Standard Model prediction~\cite{Bennett:2006fi}.

  \item {\bf Fixed target experiments.} Electron and proton beam dump
    experiments have placed limits on dark photons produced in the target and
    decaying to electron--positron pairs, see~\cite{Bjorken:2009mm,
    Batell:2009di, Essig:2010gu}. Obviously, this technique can only constrain
    dark photons with $M_{A'} > 2 m_e$. The fixed target constraints shown in
    figure~\ref{fig:paramspace} apply directly only to models \ref{model:B-L}
    and \ref{model:km}. In \ref{model:B}, $A'$ decays to electron--positron
    pairs would be loop-suppressed, so the bounds would become much weaker in
    this case. Even in model~\ref{model:km}, the constraints are not robust if
    the model contains sterile neutrinos directly charged under $U(1)'$ (the
    case of most interest to us in this work), since in this case, the
    branching ratio for $A'$ decay into $e^+ e^-$ pairs is greatly reduced.  It
    is expected that fixed target constraints will be significantly improved by
    the APEX experiment (see~\cite{Abrahamyan:2011gv} for first results from
    this experiment).

  \item {\bf $\Upsilon$ decays.} Decays of $\Upsilon$ mesons to a photon plus a
    dark photon, with the latter decaying further to $\mu^+ \mu^-$, are
    constrained by $B$-factory experiments~\cite{Essig:2009nc}.  Note that for
    a $U(1)_{B-L}$ boson these constraints are modified by an $\mathcal{O}(1)$
    factor compared to the case of a kinetically mixed dark photon because the
    electric charge of the bottom quark is different from its $B-L$ charge. In
    model~\ref{model:km}, the constraint is avoided if a fast $A'$ decay mode
    to sterile neutrinos exists.  The $U(1)_B$ model is not directly
    constrained by $\Upsilon \to \gamma \mu^+ \mu^-$, but searches for hadronic
    decays could be used to set limits, which would be strongest for $M_{A'}$ around
    the $\Upsilon$ mass $\sim 10$~GeV~\cite{Carone:1994aa, Graesser:2011vj}.
 
  \item {\bf Atomic physics constraints.} By comparing the energy differences
    between excited atomic states to the energy differences measured between
    lower-lying atomic transitions, anomalous corrections to the Coulomb force
    at atomic distance scales can be constrained~\cite{Bartlett:1988yy}. These
    constraints apply only to models in which the dark photon couples to
    electrons, i.e.\ models \ref{model:B-L} and \ref{model:km}, but not
    \ref{model:B}.

  \item {\bf Supernova 1987A (``SN1987A'').} A dark photon may be produced in the
    core of a supernova and contribute to its energy loss. By requiring that in
    Supernova 1987~A, the energy loss in dark photons was not larger than the
    known energy loss in neutrinos ($10^{53}$~erg/s), constraints on the $A'$
    mass and coupling can be derived~\cite{Dent:2012mx}. Note that these bounds
    cover only a limited range of $A'$ couplings: for too small couplings, the
    energy loss in dark photons is small; for too large couplings, the
    supernova is opaque even to dark photons, so that anomalous energy losses
    occur only in a thin outer shell. Note also that $A'$ emission in a
    supernova is mostly due to $A'$ radiation off protons and neutrons.
    Therefore, the constraints apply to all three models considered here.  Let
    us also remark that supernova constraints can be avoided altogether in
    models where the dark photon feels strong chameleon effects, so that its
    effective mass inside the supernova is higher than its mass in
    vacuum~\cite{Nelson:2007yq}.
    
  \item {\bf Solar constraints (``Sun'').} A dark photon can also be produced
    by thermal radiation in the Sun.  Unlike for supernovae, $A'$ radiation in
    stars is dominated by emission off electrons or, more precisely, by the
    conversion of plasma excitations (so-called plasmons) into dark photons.
    Requiring that the solar luminosity in dark photons is smaller than the
    known luminosity in photons places a strong bound on $A'$ bosons in a wide
    mass window around the solar plasmon resonance at $M_{A'} \sim
    100$~eV~\cite{Redondo:2008aa}. As for the supernova case, this bound is not
    robust at large $\epsilon$, where dark photon absorption inside the Sun
    becomes relevant. To our knowledge, dark photon dynamics in this regime has
    not been considered in the literature, but since it is clear that dark
    photon absorption will reduce the problematic anomalous solar energy loss
    due to $A'$ emission, part of the region labeled ``$A'$ capture in Sun'' in
    figure~\ref{fig:paramspace} may still be allowed~\cite{Redondo:2008aa}.
    Indeed, the results from~\cite{Raffelt:1988rx} (derived for keV-scale
    \emph{scalars}) suggest that $\mathcal{O}(\text{few} \times 10\
    \text{keV})$ particles may still be acceptable, while lower masses can be
    ruled out by requiring anomalous heat transfer mechanisms inside the
    Sun to be small. We expect solar constraints to be most relevant to models
    \ref{model:B-L} and \ref{model:km}, in which the dark photon couples to
    electrons. In model~\ref{model:B}, the only available $A'$ production
    mechanism is radiation off nuclei, which we expect to be much smaller than
    radiation off electrons. However, to our knowledge, this has never been
    worked out in detail.

    If the $A'$ boson couples to neutrinos, the solar constraint also depends
    on neutrino dynamics.  In the $U(1)_{B-L}$ model (model~\ref{model:B-L}),
    where the $A'$ boson couples directly to the active neutrinos, low-energy
    neutrinos produced in plasmon decay can carry away energy even if the  dark
    photon cannot because its mass is outside the plasmon resonance region. In
    the literature, the corresponding constraints are usually referred to as
    ``minicharged particle limits''~\cite{Davidson:2000hf,Jaeckel:2010ni} and
    in figure~\ref{fig:paramspace}, they are approximately indicated by the
    region labeled ``Sun/Old Stars, energy loss via $\nu$''.  At very large
    $g_{B-L}$, energy loss through low-energy neutrinos is not effective because
    these neutrinos cannot leave the Sun due to their large scattering cross
    sections. However, since neutrino scattering is suppressed by $g_{B-L}^4$,
    we estimate that the large $g_{B-L}$ values required for this to happen are
    already disfavored by other constraints, for instance atomic physics
    constraints and $g-2$ limits.  Moreover, in scenarios with very large
    $g_{B-L}$, also the spectrum of higher energy solar neutrino from nuclear
    fusion may be distorted, in potential conflict with experiments.  While a
    detailed study of these issues is beyond the scope of this work, we
    estimate that in the $U(1)_{B-L}$ model, the region labeled ``$A'$ capture
    in Sun'' in figure~\ref{fig:paramspace} is ruled out by minicharged
    particle limits.  In the $U(1)'$ model with kinetic mixing
    (model~\ref{model:km}), on the other hand, minicharged particle constraints
    are relaxed since the $A'$ in this case does not couple to active
    neutrinos, but only to hypothetical sterile neutrinos.  If the latter are
    heavier than $\sim 10$~keV, many minicharged particle limits no longer
    apply~\cite{Davidson:2000hf}. (For $\epsilon \gtrsim 10^{-9}$, even larger
    masses, $\gtrsim$~few hundred keV--several MeV may be required to avoid
    the constraints from~\cite{Davidson:2000hf}.)
    Even if the sterile neutrinos are lighter, they may
    avoid solar constraints more easily if $\epsilon$ is large. The reason is
    that for a $U(1)'$ gauge coupling of order one, the scattering of sterile
    neutrinos is suppressed only by $\epsilon^2$ (just like $A'$ absorption,
    and compared to $g_{B-L}^4$ in the $U(1)_{B-L}$ model), i.e.\ sterile
    neutrinos in this model can be more easily absorbed in the Sun.  For
    instance, for $M_{A'} = 1$~keV, $\epsilon = 10^{-5}$, the sterile neutrino
    mean free path is less than 10~cm. The average energy loss per scattering
    is of order 10~eV, i.e.\ a 1~MeV neutrino will have lost all of its kinetic
    energy after traveling for $\sim 10$~km.  We therefore estimate that the
    $U(1)'$ model with kinetic mixing, and with $M_{A'} \sim 1$~keV, $\epsilon
    \gtrsim 10^{-8}$, is not robustly excluded, even if it contains sterile
    neutrinos acting as minicharged particles. An in-depth study of the dark
    photon and sterile neutrino dynamics in the large $\epsilon$ regime is a
    possible direction for future work.

    As a final remark, we note that chameleon effects could also help
    circumvent the solar energy loss bound if the $A'$ effective mass is higher
    in the solar core than in vacuum~\cite{Nelson:2007yq, Nelson:2008tn}.

  \item {\bf Cooling of stars in globular clusters.} Evolved stars, on the horizontal
    branch in the Hertzsprung-Russell (temperature vs.\ luminosity) diagram,
    can be used to set limits on energy loss due to dark photons in the same
    way in which solar constraints are derived. A particularly promising target
    for the study of these stars are globular clusters since the initial
    conditions for star formation in these objects are the same throughout the
    cluster.  Since horizontal branch stars have a higher temperature than the
    Sun the corresponding limits reach out to higher $A'$ masses.  Like the
    constraints from solar cooling, they also disappear at large $A'$ coupling
    $\gtrsim 10^{-6}$, as approximately indicated in
    figure~\ref{fig:paramspace}. They also disappear in chameleon models,
    whereas their applicability to models without $A'$ couplings to electrons
    (model~\ref{model:B}) has not been studied in detail yet.

  \item {\bf The CAST experiment.} Helioscopes such as the CERN Axion Solar
    Telescope (CAST)~\cite{Arik:2008mq}, which are looking for electromagnetic
    signals in a dark, shielded cavity are sensitive to dark photons from the
    Sun entering the cavity and oscillating back into visible photons inside.
    This process can be used to set strong limits on the dark photon coupling
    in a mass region where the expected dark photon flux from the Sun would be
    large~\cite{Redondo:2008aa}. Helioscope bounds suffer from similar
    model-dependencies as solar and stellar constraints; in addition, they can
    be avoided if the dark photon can decays, for instance to neutrinos, before
    reaching the Earth.

  \item {\bf Light shining through walls (``LSW'').} In this type of
    experiments, one directs an intense laser beam onto an opaque wall and
    searches for residual signals behind the wall.  Conversion of laser photons
    into dark photons and back into visible photons could lead to such signals,
    and thus these experiments can be used to set limits on the dark photon
    mass and couplings~\cite{Ahlers:2007qf}.  These limits do not apply to
    $U(1)_B$ bosons (model~\ref{model:B}).

  \item {\bf CMB constraints.} A dark photon would mix with the visible photon
    in a frequency dependent way, and this effect can attenuate the black body
    spectrum of the CMB~\cite{Mirizzi:2009iz} which was measured precisely by
    the FIRAS instrument on board the COBE satellite~\cite{Fixsen:1996nj}.
    These bounds may be avoided if the mass of the dark photon is higher in the
    dense early Universe due to chameleon effects.

  \item {\bf Borexino.} Due to its large fiducial volume and the large solar
    neutrino flux, Borexino is quite sensitive to modifications of the
    neutrino--electron scattering rate. The non-observation of any anomalous
    signals can be translated into constraints on the couplings of dark photons
    to neutrinos and electrons. In particular, we require that the
    neutrino--electron scattering rate in Borexino should not be more than
    8\%~\cite{Borexino:2011rx} above the Standard Model
    prediction~\cite{Bahcall:2004pz}.  The solid red line in the top panel of
    figure~\ref{fig:paramspace} shows the Borexino constraint for the case of a
    $U(1)_{B-L}$ model (mode~\ref{model:B-L}), whereas the green lines in the
    bottom panel of figure~\ref{fig:paramspace} are for models of
    type~\ref{model:km} (a kinetically mixed $U(1)'$ gauge boson), but
    including also sterile neutrinos charged under $U(1)'$.

  \item {\bf GEMMA.} The GEMMA spectrometer at the Kalinin Nuclear Power
    Plant~\cite{Beda:2009kx} searches for anomalous contributions to
    neutrino--electron scattering, with the aim of constraining neutrino
    magnetic moments. To derive limits on the couplings of a dark photon to
    electrons, we use the data shown in figure~8 of
    reference~\cite{Beda:2010hk}\footnote{Note that the strong limit on the
    neutrino magnetic moment derived in~\cite{Beda:2010hk} was based on a
    theoretical calculation~\cite{Wong:2010pb-v1} that was later
    revised~\cite{Voloshin:2010vm, Wong:2010pb-v3}. We therefore do not
    consider this limit here, but we can still use the data presented
    in~\cite{Beda:2010hk} to derive our own limits on the parameter space of
    dark photons.} and compare it to the $A'$ model using a simple $\chi^2$
    test. We require the difference in $\chi^2$ between the $A'$ model
    and the Standard Model to be $< 3.84$, corresponding to a one-sided
    90\% C.L.\ upper limit.

    Since GEMMA is looking for neutrino--electron scattering, it is insensitive
    to the $U(1)_B$ model (model~\ref{model:B}), and also to the $U(1)'$ model
    with kinetic mixing (model~\ref{model:km}), unless there are sterile
    neutrinos charged under $U(1)'$ and heavy enough to be produced in
    oscillations of active neutrinos over distance scales of order 10~m
    (see section~\ref{sec:e-recoil-light}).

  \item {\bf Fifth force searches.} On distance scales $\gtrsim 10$~nm ($M_{A'}
    \lesssim 100$~eV), new long range forces are tightly constrained by
    precision tests of the gravitational, Casimir, and van der Waals forces.
    Since these experiments probe interactions between electrically neutral
    bodies, they are sensitive to $U(1)_{B-L}$ or $U(1)_B$ gauge
    bosons (models~\ref{model:B-L} and \ref{model:B}), but not to gauge bosons
    coupled only through kinetic mixing (model~\ref{model:km}).

    Results from fifth force searches are often reported as limits on an anomalous
    Yukawa-type contribution to the gravitational potential (see for instance
    section~24 in~\cite{Bordag:2009} or figure~28 in~\cite{Bordag:2001qi}) of
    the form $V_{\rm new} = -G_N m_1 m_2 / r \times \alpha_G \exp(-M_{A'} r)$.
    Here, $G_N$ is the gravitational constant, $m_1$, $m_2$ are the masses of
    the test bodies, $r$ is the distance between their centers of mass,
    $\alpha_G$ is a coupling constant, and $M_{A'}$ is the mass of the new
    force mediator, or equivalently the inverse range of the new force.
    Constraints on $\alpha_G$ can be translated into constraints on a $B-L$
    gauge coupling constant $g_{B-L}$ according to $g_{B-L}^2 =
    \alpha_G G_N (m_1 / q_{B-L, 1}) (m_2 / q_{B-L, 2})$, with $q_{B-L, 1}$ and
    $q_{B-L, 2}$ being the $B-L$ charges of the test bodies. In practice, $m_j
    / q_{B-L, j} \simeq 0.5$~GeV is just the nucleon mass multiplied by the
    inverse neutron fraction of the target material. (For the case of a
    $U(1)_B$ force, $q_{B-L}$ should be replaced by $q_B$, and $m_j / q_{B, j}
    \simeq 1$~GeV.)
    
    Among the most sensitive tests of the inverse square law of gravitational
    interactions over laboratory distance scales are the experiments carried
    out by the E\"ot-Wash group~\cite{Adelberger:2006dh, Adelberger:2009zz}.
    These experiments restrict the coupling strength of a $U(1)_{B-L}$ boson to
    be smaller or similar to the strength of gravitational interactions for
    $M_{A'} \lesssim 0.01$~eV (see for instance figure~1 of
    reference~\cite{Adelberger:2006dh}; for the $U(1)_B$ case, the replacement
    $g_{B-L} \to g_B \sqrt{2} N / (Z+N) \simeq 0.8g_B$ should be made in this
    figure).  Such a gauge boson could thus never lead to an observable signal
    in a dark matter detector.

    At somewhat larger masses or smaller distance scales, the strongest
    constraints come from tests of the Casimir force, i.e.\ the small
    attractive force that the zero point energy of the electromagnetic quantum
    field induces between conducting objects that are brought very close to
    each other (distance $10^{-8}$--$10^{-3}$~m)~\cite{Bordag:2001qi,
    Bordag:2009}.  At these distances, the boundary conditions imposed on the
    electromagnetic field by the conducting test bodies lead to a measurable
    modification of the electromagnetic vacuum energy. At even short distances,
    down to $10^{-9}$~m, also tests of van der Waals forces become
    important~\cite{Bordag:2001qi, Bordag:2009}.  Constraints on $g_{B-L}$ from
    tests of the Casimir effect are weaker than E\"ot-Wash limits for $M_{A'}
    \lesssim 0.01$~eV, but they extend to larger gauge boson masses and still
    provide a limit $g_{B-L} \lesssim 10^{-6}$ at $M_{A'} = 100$~eV.
\end{enumerate}

\begin{table}
  \begin{ruledtabular}
  \begin{tabular}{lccc}
                   & \parbox{4.5cm}{$U(1)_{B-L}$ (vector couplings) \\ (Model~\ref{model:B-L})}
                   & \parbox{4.5cm}{Kinetically mixed \\ (Model~\ref{model:km})}
                   & \parbox{4.5cm}{$U(1)_B$ (vector couplings) \\ (Model~\ref{model:B})} \\
    \hline
    $g-2$          & \OK & \OK                   & \NO \\
    Fixed Target   & \OK & \OK                   & \NO\footnote{By looking for hadronic
      $A'$ decays, fixed target experiments could set limits even on a $U(1)_B$
      gauge boson (model~C). To our knowledge, no such analysis has, however, been
      done.}\hspace{-0.15cm} \\
    $\Upsilon$     & \OK & \OK                   & \NO\footnote{Studies of hadronic
      $\Upsilon$ decays could be used to set limits, which would be strongest for
      $M_{A'}$ similar to the $\Upsilon$ mass, around 10~GeV~\cite{Carone:1994aa,
      Graesser:2011vj}.} \hspace{-0.15cm} \\
    Atomic physics & \OK & \OK                   & \NO \\
    Sun/Clusters/CAST & \OK & \OK                   & \NOTKNOWN \\
    SN1987A        & \OK & \OK                   & \OK \\
    LSW            & \OK & \OK                   & \NO \\
    CMB            & \OK & \OK                   & \NOTKNOWN \\
    Borexino       & \OK & only if $\nu_s$ exist & \NO \\
    GEMMA          & \OK & \NO                   & \NO \\
    Fifth force    & \OK & \NO                   & \OK \\
  \end{tabular}
  \end{ruledtabular}
  \caption{Applicability of the various constraints on new light gauge bosons
    to the three models considered here. `\OK' indicates constraints that apply to
    a given model (the exact strengths of these constraints may still differ by an
    $\mathcal{O}(1)$ factor between models), `\NO' indicates constraints
    that are not applicable, and `\NOTKNOWN' stands for constraints for which
    a dedicated study would be required to determine their applicability.}
  \label{tab:constraints}
\end{table}

\noindent
We should mention that the small allowed region between the globular cluster
and fixed target constraints can be partially ruled out in models with
\emph{only} a dark photon (but no sterile neutrinos) using the requirement that
the intergalactic diffuse photon background is not modified by $A' \to 3\gamma$
decays, that the black body spectrum of the CMB is not modified, and that the
effective number of light degrees of freedom at the epoch of Big Bang
nucleosynthesis agrees with observations~\cite{Redondo:2008ec}. The constraints
can be avoided if the dark photon decays dominantly into neutrinos (active
neutrinos in the $U(1)_{B-L}$ model and sterile neutrinos in the $U(1)'$ model
with kinetic mixing).

Moreover, for dark photons with $\mathcal{O}(\text{keV})$ masses or above, the
dark photons themselves can act as ``lukewarm dark
matter''~\cite{Redondo:2008ec}, potentially overclosing the Universe.
Constraints of this type can be avoided by the same mechanisms as for heavy
sterile neutrinos (see discussion at the end of
section~\ref{sec:kinetic-mixing}).

It is worthwhile to recapitulate which regions of parameter space are most
relevant to the phenomenology of neutrino signals in dark matter detectors. For
the kinetic mixing model (model~\ref{model:km}), dark photon masses in a window
from $\sim 10$~keV to $\sim 1$~MeV, are of most interest. Masses below 1~eV can
also be considered, provided that sterile neutrinos are sufficiently heavy
($\gtrsim 10$~keV--several hundred keV) to avoid bounds on minicharged
particles.  The kinetic mixing parameter $\epsilon$ required to produce
sizeable signals depends on the sterile neutrino flux and could vary between
$10^{-12}$ and $10^{-5}$.  For the $U(1)_{B-L}$ model (model~\ref{model:B-L},
there may be a small region of interest around $M_{A'} \sim 50$~keV, $g_{B-L}
\sim 10^{-6}$.  For other masses or couplings, chameleon effects can be
introduced to evade astrophysical and fifth force limits. The $U(1)_B$ model
(model~\ref{model:B}) is far less constrained than the other two and can yield
interesting phenomenology in the whole mass range considered (1~GeV and below).

%==============================================================================
\section{Conclusions}
\label{sec:conclusions}
%==============================================================================

In this paper, we have discussed the rich phenomenology of standard and
non-standard solar neutrino signals in dark matter direct detection
experiments. In particular, we have considered models featuring a neutrino
magnetic moment, as well as scenarios with a ``dark photon'' $A'$ (a light,
weakly coupled new gauge boson). We have shown that in these scenarios
neutrino--electron and neutrino--nucleus scattering can be much stronger than
Standard Model weak interactions at the low recoil energies to which dark
matter detectors are sensitive, while being consistent with constraints from
higher energy experiments such as Borexino and SNO. If we moreover assume that
a small fraction of solar neutrinos has oscillated into new ``sterile''
flavors, whose couplings to the dark photon are much less constrained than
those of Standard Model particles, the scattering rates can be large enough to
explain at least some of the recently reported anomalous signals from dark
matter experiments.

We have also discussed possible sources of temporal variations in the neutrino
count rate, in particular the annual variation of the Earth--Sun distance
(possibly in conjunction with oscillation effects, if the active--sterile
oscillation length is not too much smaller than one astronomical unit),
neutrino absorption in the Earth, diurnal and annual modulation due to Earth
matter effects, and temporal modulation effects in detectors whose efficiency
depends on the direction of the incoming particles, for instance because of ion
``channeling'' effects.

We conclude that dark matter detectors, with their low energy thresholds, can
be very sensitive to new physics in the neutrino sector, in particular if the
new effects are strongest at low energy and are thus hidden from dedicated
neutrino experiments, whose recoil energy thresholds are at least a few hundred
keV.  Signals from neutrino--electron scattering and neutrino--nucleus
scattering can be easily confused with dark matter scattering, especially when
the signal consists of only a few events so that detailed investigations of the
event spectrum or of temporal modulation effects are not possible.  This is an
important consideration for the present generation of detectors, but even more
so for future experiments, whose large mass and reduced background will make
them more sensitive to both dark matter and neutrino signals.  On the other
hand, our study also shows that dark matter detectors are powerful tools to
constrain or discover neutrino physics beyond the Standard Model.

{\it Note added:} Shortly after the present paper was completed, an interesting
related work by Pospelov and Pradler appeared on the
arXiv~\cite{Pospelov:2012gm}.

\begin{acknowledgments}
The authors would like to thank Rouven Essig, Francesc Ferrer, Kai
Schmidt-Hoberg, Maxim Pospelov, Surjeet Rajendran, Carlos Savoy, Ian Shoemaker,
Neil Weiner, our anonymous referee, and especially Toshihiko Ota and Javier
Redondo for interesting and useful discussions. JK is grateful to the Aspen
Center for Physics (supported by the National Science Foundation under Grant
No.~1066293), where part of this work has been carried out.  PANM would like to
thank Fermilab for kind hospitality and support during his visits. PANM is
supported by the Funda\c{c}\~{a}o de Amparo \`{a} Pesquisa do Estado de S\~{a}o
Paulo and by the European Commission under contract PITN-GA-2009-237920.
Fermilab is operated by Fermi Research Alliance, LLC, under contract
DE-AC02-07CH11359 with the United States Department of Energy.
\end{acknowledgments}

%==============================================================================
\appendix
\section{Models with heavy sterile neutrinos}
\label{sec:heavy-nus-model}
%==============================================================================

In section~\ref{sec:electron-recoil} we have considered heavy sterile neutrino
with masses of order several hundred keV, which led to interesting features in
the recoil energy spectrum (see figure~\ref{fig:Zprime-heavy-sterile}) and
helped us avoid constraints on neutrino--electron scattering from XENON-100.
On the other hand, in order to get a large enough scattering rate at low
energies, the dark photon which mediates the interactions between sterile
neutrinos and electrons has to be fairly light, at or below $\sim 10$~keV. Furthermore
the heavy sterile neutrino needs to mix with the light active neutrinos to be
produced in the Sun. The product of couplings and mixing angles $g_e g_\nu
\sin\theta$ should be of order $10^{-11}$, i.e.\ with $g_\nu \lesssim 1$ from
perturbativity and $g_e \lesssim 10^{-5}$ (see figure~\ref{fig:paramspace}),
this translates into the requirement $\sin\theta \gtrsim 10^{-6}$

Achieving this within a see-saw model similar to the one introduced in
section~\ref{sec:model-intro} (equations~\eqref{eq:L-light-steriles} and
\eqref{eq:seesaw}) is not possible because in such models the mass of the
sterile neutrino is of order $\ev{H'}^2/M_R\ll \ev{H'},\,M_{A'}$ when the
right-handed neutrinos are very heavy.

Apart from the neutrino masses, we also need to generate an appropriate mixing
matrix. In particular, in order for the sharp drop in the spectra visible in
figure~\ref{fig:Zprime-heavy-sterile} to occur, the electron recoil signal has
to be dominated by scattering of the heavy mass eigenstate $\nu_4$ rather than
by scattering of the light mass eigenstates $\nu_1$, $\nu_2$, $\nu_3$ through
their small $\nu_s$ admixture. (Remember that it is the the sterile
\emph{flavor} eigenstate $\nu_s$ which is charged under the new gauge group.)
This requires the leptonic mixing matrix element $U_{e4}$ (which gives the
$\nu_4$ admixture to the electron neutrino) to be much larger than the elements
$U_{s1}$, $U_{s2}$, $U_{s3}$ (which give the $\nu_s$ component of the light
mass eigenstates).

Building a model that satisfies all these requirements requires some
engineering, but is certainly not impossible, as we now show.  We add a pair of
sterile neutrinos, one of them (say, $\nu_{sL}$) charged under $U(1)'$ and the
other one ($\nu_{sR}^c$) neutral (anomalies can be taken care of by introducing
extra spectator fermions).  The Dirac mass of the sterile neutrinos is thus
proportional to $\ev{H'}$, which we will take to be around 800~keV.  The mass
of the dark photon is also proportional to $\ev{H'}$, but may be smaller if the
$U(1)'$ gauge coupling is smaller than one. For instance, let us assume
a gauge coupling of order 0.01~to give the dark photon a mass around 10~keV.
The presence of the $H'$ vev also allows us to write mixing terms between
active and sterile neutrinos, particularly between the left handed sterile and
the right handed active states. The full neutrino mass matrix is
\begin{align}
  \renewcommand{\arraystretch}{1.3}
  -\mathcal{L}_m \supset \overline{\psi_\nu^c}
    \left( \mbox{   \begin{tabular}{c@{\quad}c@{\ \ }|@{\ \ }c@{\quad}c}
      0                    & $Y_\nu^* \ev{H}$ & 0                & $Y_{\nu4}^* \ev{H}$ \\
      $Y_\nu^T \ev{H}$     & $M_R$            & $Y_s'^* \ev{H'}$ & $M_{R,14}$ \\[0.1cm]
      \hline
      0                    & $Y_s'^T \ev{H'}$ & 0                & $Y_s^* \ev{H'}$ \\
      $Y_{\nu 4}^T \ev{H}$ & $M_{R,14}^T$     & $Y_s^T \ev{H'}$  & $M_{R,44}$
    \end{tabular}} \right)  \, \psi_\nu \,.
  \label{eq:mnu-heavy}
\end{align}
where we have placed all neutrinos in a vector $\psi_\nu = [\nu_{L},
(\nu_{R})^c,\, \nu_{sL},\, (\nu_{sR})^c]$, and flavor indices are suppressed.
If there are, in addition to the three active neutrino flavors and their
right-handed partners, $n_s$ left-handed sterile neutrinos and the same number
of right-handed sterile neutrinos, $Y_\nu$ and $M_R$ are understood to be $3
\times 3$ matrices, $Y_{\nu 4}$, $Y_s$, and $M_{R,14}$ are $3 \times n_s$
matrices, and $Y_s$, $M_{R,44}$ are $n_s \times n_s$ matrices.

To obtain a model with the desired masses and mixing angles, we will assume
that $Y_{\nu 4}$, $M_{R,14}$, and $M_{R,44}$ are negligibly small. This could,
for instance, be justified by declaring sterile lepton number (carried by $\nu_{sL}$
and $\nu_{sR}$) a global symmetry only broken by the spurion $Y'_s$.
Another possibility is to introduce \emph{two} $U(1)'$-breaking Higgs fields,
$H'_1$ and $H'_2$, with only $H'_1$ carrying sterile lepton number. Then,
the term $Y_s'^* \ev{H'}$ in equation~\eqref{eq:mnu-heavy} would be replaced
by $Y_s'^* \ev{H'_1}$, and the term $Y_s^* \ev{H'}$ would be replaced by
$Y_s^* \ev{H'_2}$. The terms $M_{R,14}$, $M_{R,44}$, and $Y_{\nu4}^* \ev{H}$,
on the other hand, would be absent.

If we set $Y_{\nu 4}$, $M_{R,14}$, and $M_{R,44}$ to zero and assume
$M_R \gg \ev{H}, \ev{H'}$ in equation~\eqref{eq:mnu-heavy}, we find that
to leading order in the small masses, the admixture of the $\mathcal{O}(\ev{H'})$
mass eigenstates to the left-handed active flavor eigenstates is
\begin{equation}
  |U_{e4}| \sim \frac{Y_\nu Y'_s}{\sqrt{2} Y _s}\,\frac{\ev{H}}{ M_R }\,.
  \label{eq:mixing-heavy-model}
\end{equation}
The admixture of the $U(1)'$-charged flavor eigenstate $\nu_{sL}$ to the
light $\mathcal{O}(\ev{H}^2 / M_R)$ mass eigenstates, on the other hand,
vanishes to leading order as desired.

As is clear from equation~\eqref{eq:mixing-heavy-model}, a standard seesaw
mechanism with $M_R \sim 10^{15}$~GeV will produce mixing angels which are too
small.  However an electroweak-scale seesaw with $M_R \sim 100$~GeV can easily
yield mixings that are large enough. For example, taking $Y_s$ and $Y'_s$ of
order one, and requiring the active neutrino masses $m_\nu \sim (Y_\nu \ev{H})
M_R^{-1} (Y_\nu \ev{H})$ to be at the 0.1~eV scale, gives a mixing angle of
\begin{equation}
  |U_{e4}| \sim 10^{-6}\left(\frac{\text{100 GeV}}{M_R}\right) \,.
  \label{eq:mixing-example}
\end{equation}
In order to produce spectra that are close to those observed by DAMA and CoGeNT
one can pick $g_\nu\sim 1$ and $g_e \sim 10^{-5}$, which is not ruled out by
the constraints from section~\ref{sec:paramspace}.

%==============================================================================
\section{Earth matter effects on solar neutrinos}
\label{sec:nuosc}
%==============================================================================

In this appendix we explain in detail how Earth matter effects in the sterile
neutrino sector can lead to daily modulation (see
section~\ref{sec:earth-matter-modulation})~\cite{Akhmedov:2004rq}. We assume a
model with two sterile flavor eigenstates $\nu_{s1}$ and $\nu_{s2}$, with
$\nu_{s1}$ giving the strongest signal in the detector. Thus, we are interested
in the probability for $\nu_{e}\to\nu_{s1}$ transitions. As in
section~\ref{sec:earth-matter-modulation}, we make the simplifying assumption
that the new, mostly sterile, mass eigenstates $\nu_4$ and $\nu_5$ are so heavy
that they cannot be produced in a coherent superposition with the light mass
eigenstates $\nu_1$, $\nu_2$, $\nu_3$. We do assume, however, that coherence
between $\nu_4$ and $\nu_5$ is possible due to a small mass splitting. As the
neutrinos are produced in the core of the Sun, and as we assume the oscillation
length to be much smaller than the size of the Sun's core, the solar neutrino
flux at the Earth can be treated as an \emph{incoherent} mixture of $\nu_4$
and $\nu_5$.  Hence, the computation of the probability $\nu_{e}\to\nu_{s1}$ is
two-step: first, we need to determine the amount of $\nu_{4,5}$ that exit the
Sun; then we need to calculate the probability of detecting these mass
eigenstates as $\nu_{s1}$ after they have traveled a distance $L^\oplus$ in
the Earth. Finally, we perform an incoherent sum, that is we compute
\begin{align}
  P(\nu_{e} \to \nu_{s1}) = P_{\odot}(\nu_{e}\to\nu_{4}) P_{\oplus}(\nu_{4}\to\nu_{s1})
    + P_{\odot}(\nu_{e}\to\nu_{5})P_{\oplus}(\nu_{5}\to\nu_{s1}),
  \label{eq:Pnuenus1}
\end{align}
where $P_{\odot}$ and $P_{\oplus}$ are the transition probabilities in the Sun
and in the Earth, respectively.

Since only the two flavors $\nu_{s1}$ and $\nu_{s2}$ are relevant to us,
we can define an effective two-flavor rotation matrix with mixing angle $\theta$
\begin{align}
  U^\theta = \left(\begin{array}{cc} \cos\theta &
              \sin\theta\\ -\sin\theta & \cos\theta\end{array}\right) \,.
  \label{eq:mix-matrix-2f}
\end{align}
We denote the vectors of the vacuum mass basis, flavor basis and the effective
mass basis in Earth matter
by $|\nu_{i}\rangle$, $|\nu_{\alpha}\rangle$, and
$\ket{\nu_{i}^\oplus}$, respectively. The relations between them are
given by
\begin{align}
  |\nu_{\alpha}\rangle = U_{\alpha i}^{\theta *} |\nu_{i}\rangle
  \qquad\text{and}\qquad
  |\nu_{\alpha}\rangle = U_{\alpha i}^{\oplus *}|\nu_{i}^\oplus\rangle \,,
  \label{eq:eigenstates}
\end{align}
where $U^\oplus$ is the effective mixing matrix in Earth matter, which has the
same form as equation~\eqref{eq:mix-matrix-2f}, but with $\theta$ replaced by
the effective mixing angle in Earth matter, $\theta_\oplus$, which is obtained
when the full Hamiltonian is diagonalized.  Concretely, working in the flavor
basis we have
\begin{align}
  H = \frac{1}{4E} U^\theta
      \left(\begin{array}{cc} -\Delta m^{2} & 0\\ 0 & \Delta m^{2}\end{array}\right)
      U^{\theta\dagger}
      + \left(\begin{array}{cc} V_{A'}^\oplus & 0\\ 0 & 0\end{array}\right) \,,
\end{align}
(with $\Delta m^2 \equiv m_5^2 - m_4^2$) and
\begin{align}
  U^{\oplus\dagger} \, H \, U^\oplus \equiv
  \text{diag}(-\Delta m_\oplus^{2}/4E \ ,\ \Delta m_\oplus^{2}/4E) \,.
\end{align}
It is easy to show that 
\begin{align}
  \cos2\theta^\oplus = \frac{\Delta m^{2} \cos2\theta - 2EV_{A'}^\oplus}
                            {\Delta m_\oplus^{2}}
  \qquad\qquad\text{and}\qquad\qquad
  \sin2\theta^\oplus = \frac{\Delta m^{2}\sin2\theta}{\Delta m_\oplus^{2}} \,,
\end{align}
where
\begin{align}
  \Delta m_\oplus^{2} = \omega \, \Delta m^{2}
                      = \Delta m^2 \sqrt{\sin^{2}2\theta
                            + (\cos^{2}2\theta-2EV_{A'}^\oplus/\Delta m^{2})^{2}} \,.
\end{align}
Now, we calculate $P_{\oplus}(\nu_{4}\to\nu_{s1}) = |\bra{\nu_{s1}}
e^{-i H L_\oplus} \ket{\nu_4}|^{2}$. With the abbreviations $c_\theta \equiv \cos\theta$,
$s_\theta \equiv \sin\theta$, $c_\oplus \equiv \cos\theta_\oplus$ and
$s_\oplus \equiv \sin\theta_\oplus$, we find
\begin{align}
  \big|\bra{\nu_{s1}} e^{-i H L_\oplus} \ket{\nu_4}\big|^2
    &= \Bigg|\begin{pmatrix} 1 & 0 \end{pmatrix}
       \begin{pmatrix}
         c_\oplus & s_\oplus \\
        -s_\oplus & c_\oplus
       \end{pmatrix}
       \begin{pmatrix}
         e^{+i\Delta m_\oplus^{2}L_\oplus/4E} & 0 \\
         0                                    & e^{-i\Delta m_\oplus^{2}L_\oplus/4E}
       \end{pmatrix}
       \begin{pmatrix}
         c_\oplus & -s_\oplus \\
         s_\oplus &  c_\oplus
       \end{pmatrix}
       \begin{pmatrix}
         c_\theta \\ -s_\theta
       \end{pmatrix} \Bigg|^2 \\
    &= \cos^2\theta - \sin^2 2\theta \frac{2 E V_{A'}^\oplus}{\omega^2 \Delta m^2}
                      \sin^2 \bigg( \frac{\omega \Delta m^{2} L_\oplus}{4E}\bigg)
  \label{eq:Pearth}
\end{align}
In the first line we have used the relations~\eqref{eq:eigenstates}, and the
third line then follows from simple algebraic manipulations.  It is
straightforward to see that to obtain $P_{\oplus}(\nu_{5}\to\nu_{s1})$ we just
need to replace $\cos^{2}\theta$ by $\sin^{2}\theta$ and reverse the sign of
the second term.

To complete the calculation of $P(\nu_e \to \nu_{s1})$, we need
$P_{\odot}(\nu_e \to \nu_{4,5})$. The probability for producing the matter
eigenstates corresponding to $\nu_{4,5}$ at the center of the Sun
is given by $|U_{e4,5}^\odot|^2$, and since the adiabaticity condition
\eqref{eq:adiabaticity} is well fulfilled (unless $\sin 2\theta$ is tiny, in
which case we would not expect an observable signal from sterile neutrino
scattering anyway), this is also the probability that a solar neutrino
exits the Sun as $\nu_{4,5}$, i.e.\
\begin{align}
  P_{\odot}(\nu_e \to \nu_{4,5}) = |U_{e4,5}^\odot|^2 \,.
  \label{eq:Psun}
\end{align}
Plugging \eqref{eq:Psun} and \eqref{eq:Pearth} into \eqref{eq:Pnuenus1},
we obtain equation~\eqref{eq:solar-prob}.

% Bibliography
\bibliographystyle{apsrev}
\bibliography{./nu-dama}

\end{document}